\documentclass[useAMS,usenatbib]{mnras}
\usepackage{graphics} 
\usepackage{epsfig, color} 
\usepackage{graphicx}
\usepackage{amsmath,ulem} 
\usepackage[T1]{fontenc}

\def\ltsima{$\; \buildrel < \over \sim \;$}
\def\simlt{\lower.5ex\hbox{\ltsima}}
\def\gtsima{$\; \buildrel > \over\sim\;$}
\def\simgt{\lower.5ex\hbox{\gtsima}}

\def\msun{{\,{\rm M}_\odot}}

\def\del#1{{}}

\title[AGN outflows and star formation]{Do AGN outflows quench or enhance star formation?}  
\author[]{Kastytis Zubovas$^{1,2\star}$ and Martin A. Bourne$^{3}$ \\
$^{1}$Center for Physical Sciences and Technology, Savanori\c{u} 231, Vilnius LT-02300, Lithuania\\
$^{2}$Astronomy Observatory, Faculty of Physics, Vilnius University, M. K. \v{C}iurlionio 29, Vilnius LT-03100, Lithuania\\
$^{3}$Institute of Astronomy and Kavli Institute for Cosmology, University of Cambridge, Madingley Road, Cambridge, CB3 0HA, UK\\
$^{\star}$ {E-mail:~} {\rm kastytis.zubovas@ftmc.lt} } 

\begin{document} %\date{Received} \pagerange{\pageref{firstpage}--\pageref{lastpage}}
%\pubyear{2016} \maketitle \label{firstpage} 

\maketitle 

\begin{abstract} 

AGN outflows can remove large quantities of gas from their host galaxy
spheroids, potentially shutting off star formation. On the other hand,
they can compress this gas, potentially enhancing or triggering star
formation, at least for short periods. We present a set of idealised
simulations of AGN outflows affecting turbulent gas spheres, and
investigate the effect of the outflow and the AGN radiation field upon
gas fragmentation. We show that AGN outflows of sufficient luminosity
shut off fragmentation while the nucleus is active, but gas
compression results in a burst of fragmentation after the AGN switches
off. Self-shielding of gas against the AGN radiation field allows some
fragmentation to occur during outbursts, but too much shielding
results in a lower overall fragmentation rate. For our idealised
simulation setup, there is a critical AGN luminosity which results in
the highest fragmentation rate, with outflows being too efficient at
removing gas when $L > L_{\rm crit}$ and not efficient enough to
compress the gas to high densities otherwise. These results, although
preliminary, suggest that the interaction between AGN and star
formation in their host galaxies is particularly complex and requires
careful study in order to interpret observations correctly.

\end{abstract}


\section{Introduction}

Over the last two decades, essentially irrefutable evidence has been
gathered that all massive galaxies \citep[e.g.,][]{Kormendy2013ARA&A},
and perhaps even some dwarf galaxies \citep{Lemons2015ApJ} contain
supermassive black holes (SMBHs) in their centres and that the masses
of these SMBHs correlate with various galaxy parameters, such as the
stellar velocity dispersion in the spheroidal component \citep[the
  $M_{\rm BH}-\sigma$ relation;][]{Ferrarese2000ApJ, Gultekin2009ApJ,
  McConnell2013ApJ} or the stellar mass of the spheroid \citep[the
  $M_{\rm BH}-M_{\rm b}$ relation;][]{Haering2004ApJ,
  McConnell2013ApJ, Kormendy2013ARA&A}. The usual interpretation of
these relations is that they are a signature of co-evolution between
SMBHs and their hosts over cosmological timescales
\citep{Kauffmann2000MNRAS, DiMatteo2005Natur}, although this is not a
universally accepted explanation \citep[e.g.,][]{Hirschmann2010MNRAS};
for a review, see \citet{Kormendy2013ARA&A}. A way for the SMBH to
impact the properties of the host galaxy is via large-scale outflows
driven during the AGN phases; such outflows have been observed in
numerous nearby galaxies \citep{Feruglio2010A&A, Sturm2011ApJ,
  Cicone2014A&A}. We note, however, that large-scale outflows are not
the only possible explanation of co-evolution. See, e.g.,
\citet{Sazonov2005MNRAS} for a radiative-feedback-based model and
\citet{Angles2015ApJ, Angles2017MNRAS} for a model based on
gravitational torque-driven SMBH accretion. It is even possible that
the correlations appear due to statistical effects of galaxy mergers,
rather than any causal connection \citep{Jahnke2011ApJ}.

The mass, energy and momentum fluxes of these outflows are well
explained by the AGN wind-driven feedback model \citep{King2003ApJ,
  King2010MNRASa, Zubovas2012ApJ}. The primary mechanism responsible
for inflating the outflows is radiation pressure from the AGN
accretion disc \citep[although MHD-driven winds have also been
  suggested, see][]{Fukumura2015ApJ}, which launches a
quasi-relativistic ($v \sim 0.1c$) wide-angle wind, carrying $\sim
5\%$ of the AGN luminosity as kinetic power. The wind shocks against
the ambient medium and heats up to very high temperatures. Theoretical
arguments \citep{Faucher2012MNRASb} and lack of observational evidence
of cooling wind shocks \citep{Bourne2013MNRAS} suggest that under
realistic circumstances, the shocked wind bubble cools inefficiently
and hence transfers most of its energy to the surrounding medium. This
results in the formation of an energy-driven outflow, which expands
with a velocity of order $1000$~km~s$^{-1}$ and can have a mass flow
rate of $>1000 \msun$~yr$^{-1}$. The momentum flux of the outflow is
more than an order of magnitude higher than the momentum flux of the
AGN radiation field $L_{\rm AGN}/c$ \citep{Zubovas2012ApJ};
\citet{Faucher2012MNRAS} and \citet{Stern2016ApJ} predict similarly
large momentum-loading factors.

The outflow has the potential to clear out gas from the spheroid of
its host galaxy, thus quenching further star formation
\citep{Hopkins2010MNRAS, Zubovas2012ApJ}. On the other hand, the
outflow might compress dense gas clouds as it flows around them,
potentially enhancing star formation \citep{Silk2005MNRAS,
  Silk2013ApJ, Ishibashi2012MNRAS, Nayakshin2012MNRASb,
  Zubovas2014MNRASc}. It has been suggested that AGN-induced star
formation may be responsible for ultraluminous starbursts
\citep{Silk2005MNRAS} and the size growth of galaxies from $z \sim 2$
to present day \citep{Ishibashi2013MNRAS, Ishibashi2014MNRAS}. Other
models of AGN-induced star formation have been proposed as well, such
as the interaction of an AGN jet with the interstellar medium
\citep{Silk2012A&A, Gaibler2012MNRAS} or the shocks and compression
induced by the backflow of a jet-inflated bubble on to the galactic
disc \citep{Bieri2016MNRAS}. There is some observational evidence for
the triggering of star formation by jet coccoons
\citep{Crockett2012MNRAS}. It is currently not clear which of the two
processes - gas removal or triggering of star formation - is more
important. Observational evidence does not clarify this picture
either: starbursts are sometimes observed to occur before the onset of
nuclear activity in galaxies \citep[e.g.,][]{Schawinski2009ApJ}, but
an opposite situation, that is, a starburst caused by an AGN, would be
impossible to distinguish due to the short lifetimes of AGN
\citep{Schawinski2015MNRAS, King2015MNRAS}. \citet{Silk2010ApJ} even
suggest, based on analytical calculations, that AGN-induced star
formation is necessary for gas removal from galaxies by outflows,
further complicating the issue.

In this paper, we present results of a number of idealised simulations
designed to investigate the effect of AGN outflows on the
fragmentation of a turbulent gas distribution. With these, we set out
to answer two questions:

\begin{itemize}

\item What is the effect of outflows driven by AGN of various
  luminosities upon the spatially-integrated fragmentation rate in a
  turbulent gas distribution?

\item What properties does the gas that eventually turns into stars
  have just before the passage of the outflow? In other words, what
  kind of gas is susceptible to fragmentation due to (or despite) the
  passage of the AGN outflow, and what kind of gas is removed from the
  galaxy instead?
  
\end{itemize}

We find that during the period of activity, AGN radiation and outflows
typically quench or at least limit fragmentation rates, but
fragmentation resumes after the AGN switches off. There is a critical
AGN luminosity which leads to the highest average post-AGN star
formation rate. Sink particles tend to form either in dense infalling
filaments or at the edges of outflow bubbles. In the latter case, the
sink particles have systematically higher radial velocities than gas
does on average, which could lead to corresponding stars being
observationally distinguishable from stars formed elsewhere in the
galaxy. On smaller scales, most of the pre-existing dense clumps are
obliterated by AGN outflows and the material that fragments after the
AGN switches off is different than the material that would have
fragmented without AGN influence.

We structure the paper as follows. In Section \ref{sec:sims}, we
present the numerical model of our simulations. The results are shown
in Section \ref{sec:results}. Finally, we discuss and summarize these
results in Section \ref{sec:discuss}, with particular emphasis on how
they can inform more detailed simulations and analysis of
observations.

\section{Simulation setup} \label{sec:sims}

The simulation set-up is similar to that used in our previous
simulations of AGN outflows \citep[e.g.,][]{Nayakshin2012MNRASb,
  Bourne2015MNRAS, Zubovas2016MNRASa}. We use the hybrid SPH/N-body
code \textsc{GADGET-3} \citep[a modified version of the publically
  available \textsc{GADGET-2}][]{Springel2005MNRAS}, with the SPHS
\citep{Read2010MNRAS, Read2012MNRAS} flavour of SPH and the Wendland
kernel \citep{Wendland95,Dehnen2012MNRAS} with 100 neighbours.

The initial conditions for all simulations are a spherical shell of
gas with inner and outer radii $R_{\rm in} =0.1$ and $R_{\rm out} = 2$
kpc, respectively. The shell has a $\rho \propto R^{-2}$ density
profile and a total mass of $M_{\rm sh} = 6.1 \times 10^9 \msun$,
tracked with $N = 10^6$ particles, giving a particle mass of $m_{\rm
  SPH} = 6100 \msun$. The shell is placed in a background isothermal
potential with a 1D velocity dispersion $\sigma = 154$ km s$^{-1}$ and
is initially given a turbulent velocity field, with a characteristic
velocity $v_{\rm turb} = \sigma$. We allow the gas to evolve for
$1$~Myr before turning the AGN on (see below), in order for the
density field to become inhomogeneous. During this relaxation period,
the gas is affected by all relevant processes - its own gravity, the
gravity of the background potential, hydrodynamic forces, including
the initial turbulent velocity field, as well as cooling using the
same \citet{Sazonov2005MNRAS} prescription as is done during the rest
of the simulation (see below).

After the initial relaxation period, the AGN is turned on in the
centre of the gas distribution. The AGN luminosity $L_{\rm AGN}$ is
fixed to a constant value for the whole duration of its activity
$t_{\rm q}$; $L_{\rm AGN}$ and $t_{\rm q}$ are the free parameters in
our simulations. The AGN affects gas by heating it directly, and by
producing wind feedback. Gas heating and cooling is treated using the
prescription from \citet{Sazonov2005MNRAS}, which is specifically
designed to follow optically thin gas exposed to a standard AGN
spectrum. This cooling function was modified to neglect Compton
cooling, a change appropriate for the two-temperature plasma, which is
the state of the hottest outflowing material. The prescription does
not include the effects of radiation pressure. In principle, this is a
drawback, however in our particular case, the energy-driven AGN
outflow provides a momentum rate at least one order of magnitude
greater than the momentum rate (pressure force) of the radiation
field, therefore we think it is not unreasonable to neglect the effect
of radiation pressure.

In addition, in several
simulations we implement a crude approximation of gas optical depth,
by calculating the radial profile of optical depth of the initial gas
distribution, and multiplying that value by the local gas density:
\begin{equation}\label{eq:optdepth}
\tau = \kappa f_{\rm shield} \rho R\left(\frac{R}{R_{\rm in}}-1\right).
\end{equation}
Here, $\kappa = 0.346$~cm$^2$~g$^{-1}$ is the electron scattering
opacity, $f_{\rm shield}$ is a dimensionless free parameter, $\rho$ and $R$
are the density and radial coordinate of the gas particle, and $R_{\rm
  in} = 0.1$~kpc is the inner edge of the initial gas
distribution. This approximation implicitly assumes that all
significant deviations from the initial density distribution happen in
the vicinity of the particle in question. This is not accurate, but
does not require almost any extra computation time, while providing
some shielding from the effects of AGN radiation in the densest
regions and allowing them to cool and fragment more efficiently than
otherwise.

The \citet{Sazonov2005MNRAS} cooling function is valid only for
temperatures above $10^4$~K; for colder gas, we adopt the cooling
function of \citet{Mashchenko2008Sci}, which allows cooling down to
$20$~K, incorporating atomic, molecular and dust-mediated cooling in
Solar-metallicity gas.

AGN wind feedback is modelled with the virtual particle method of
\citet{Nayakshin2009MNRASb}. The AGN emits tracer particles
spherically symmetrically; these particles represent the AGN wind,
which is too dilute and too fast for full hydrodynamical modelling to
be feasible in SPH simulations. The virtual particles carry momentum
$p_{\rm v} = 0.1 m_{\rm SPH} \sigma$ and energy $E_{\rm v} = 0.05
p_{\rm v}c$ in straight lines, and transfer these quantities to SPH
particles encountered along the way. The transfer of momentum and
energy represents the interaction between the wind and the surrounding
ISM. This method does not produce a reverse wind shock, however the
contact discontinuity and the forward shock are produced and evolve
consistently with the analytical predictions \citep{Zubovas2012MNRASa,
  Zubovas2012ApJ}. Since we consider situations where the wind shock
cools inefficiently, we allow all of the virtual particle energy to be
transferred to the SPH particles.  This method, by construction, takes
into account different optical depths along different lines of sight
from the AGN and allows simultaneous gas inflows and outflows to form
in a multiphase medium \citep{Zubovas2016MNRASa}.

The mass resolution of our simulations does not allow us to track the
formation of individual stars. Instead, we adopt a density-based
temperature floor, which ensures that the Jeans mass of all gas
particles does not fall below the resolution limit:
\begin{equation} 
\begin{aligned}
 T_{\rm floor} & = \rho^{1/3}\frac{\mu m_{\rm P}G}{\pi k_{\rm B}}\left(N_{\rm ngb}m_{\rm SPH}\right)^{2/3} \\  
   & \simeq 850\left(\frac{\rho}{10^{-22}{\rm g cm}^{-3}}\right)^{1/3}\left(\frac{\mu}{0.63}\right)\left(\frac{m_{\rm sph}}{6100M_{\rm \odot}}\right)^{2/3}K
\end{aligned} 
\label{T_floor} 
\end{equation} 
where $\mu=0.63$ is the mean molecular weight and $N_{\rm ngb}=100$ is
the number of SPH neighbours. Particles that lie on this temperature
floor can turn into sink particles with a probability
$P=1-\exp(-0.1\Delta t/\tau_{\rm ff})$, where $\Delta t$ is the
current particle time-step and $\tau_{\rm ff}$ is the local free-fall
time. The sink particles represent dense molecular clouds, which we
expect to form stars rather quickly. This expression assumes that the
efficiency of fragmentation is $10\%$, i.e. $10\%$ of the gas that can
fragment will do so in one dynamical time. The actual star formation
efficiency (SFE) is not straightforward to determine. In an
unperturbed ISM, for structures of the mass scale equivalent to the
particle masses in our simulations, it is of order a few percent
\citep{McKee2007ARA&A}, so our estimates of fragmentation rate could
be taken as an upper limit to the star formation rate. On the other
hand, external compression can enhance the star formation efficiency
\citep{Whitworth1994A&A, Zubovas2014MNRASc}, so the actual SFE might
be greater than quoted above. Stellar feedback complicates the issue
further, as it can both reduce the density of the surrounding gas and
hence the fragmentation rate \citep{Bate2009MNRAS} or compress
surrounding material and enhance fragmentation; neither process is
modelled in our simulations. Therefore we caution the reader to not
interpret the fragmentation rates given in the Results section as
equivalent to the star formation rates, however the qualitative
behaviour and comparison of different systems should still be valid
even with this approximation in place.

\section{Results} \label{sec:results}

\subsection{Simulation parameters}

There are two free parameters that we vary in most of our
simulations. The AGN luminosity is given one of seven values: $L_{\rm
  AGN} = L \times 1.3 \times 10^{46}$~erg~s$^{-1}$, where $L = 0.5, 1,
2, 3, 4, 5$ or $10$. The luminosity value with $L = 1$ corresponds to
the Eddington luminosity for a $10^8 \msun$ SMBH. It is also the
critical luminosity required for the AGN wind momentum to clear the
gas from a $\sigma = 154$~km~s$^{-1}$ potential, assuming that the gas
is smoothly distributed and the outflow propagates spherically
symmetrically. As we have shown in previous papers
\citep{Bourne2015MNRAS, Zubovas2016MNRASa} and show again below, this
luminosity is not high enough to drive an outflow in a turbulent
medium.

The duration of the AGN episode is either constrained to $t_{\rm q} =
1$~Myr, or kept infinite. Although a $1$~Myr duration is longer than
the timescale over which AGN are expected to flicker \citep[$t_{\rm
    fl} \simeq 5\times 10^4$~yr,][]{Schawinski2015MNRAS,
  King2015MNRAS}, we expect that multiple flickering episodes combine
to produce a single outflow. $1$~Myr might be a reasonable duration
for a single AGN feeding event, i.e. the time required for a
several-parsec-scale gas reservoir to be consumed. Thus the two cases
of $t_{\rm q}$ allow us to investigate how the surrounding gas reacts
both to a driven AGN outflow and to a coasting one.

We also run three simulations investigating the effect of gas
self-shielding, as described above. These simulations are based on the
one with the luminosity value $L = 5$ and $t_{\rm q} = 1$~Myr and differ by
the adopted self-shielfing factor $f_{\rm shield}$; we use $f_{\rm
  shield} = 1, 10, 1000$ to investigate these differences.

Each simulation is labelled using the following convention:
L\#[T1[Sz]], where $\#$ is the value $L$ of luminosity (see above), T1
identifies simulations with $t_{\rm q} = 1$~Myr and the number next to
S gives the value of $f_{\rm shield}$; simulations without a T1 identifier
have $t_{\rm q} = \infty$, while simulations without an Sz identifier
have $f_{\rm shield} = 0$.

\subsection{Outflow propagation}

\begin{figure*}
    \centering
    \includegraphics[trim = 6mm 27mm 6mm 9mm, clip, width=0.3 \textwidth]{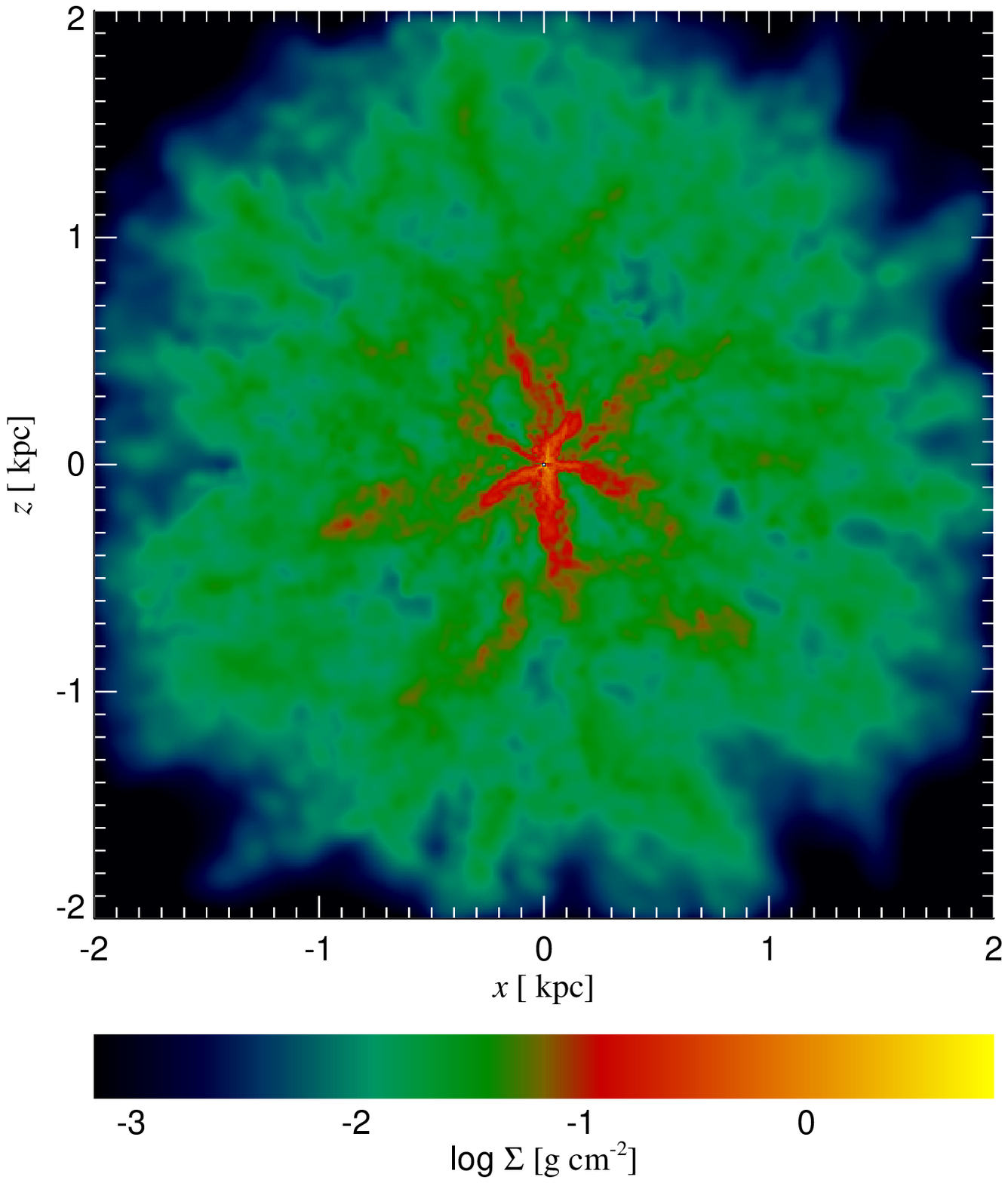}
    \includegraphics[trim = 6mm 27mm 6mm 9mm, clip, width=0.3 \textwidth]{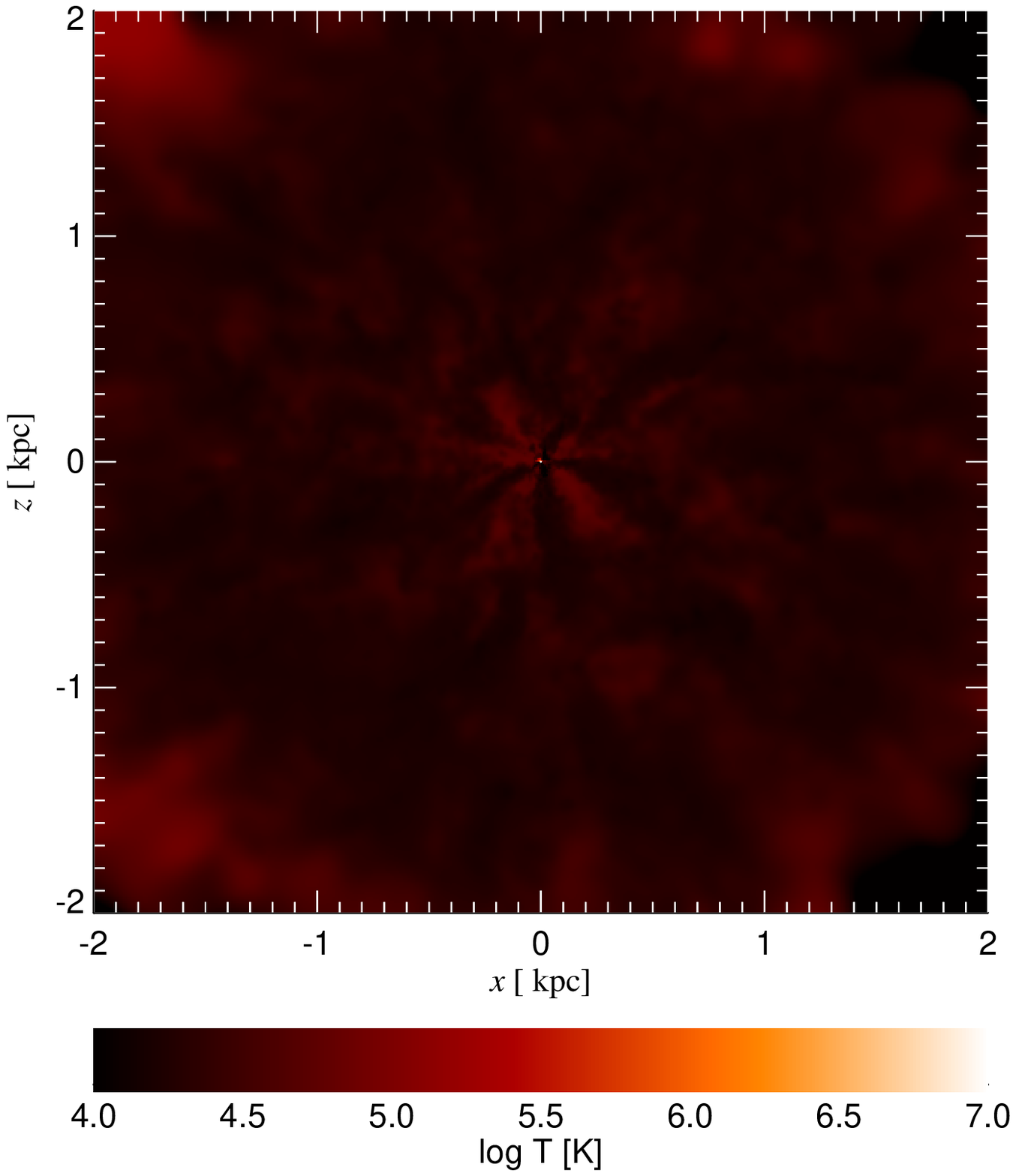}
    \includegraphics[trim = 0 0 0 0, clip, width=0.33 \textwidth]{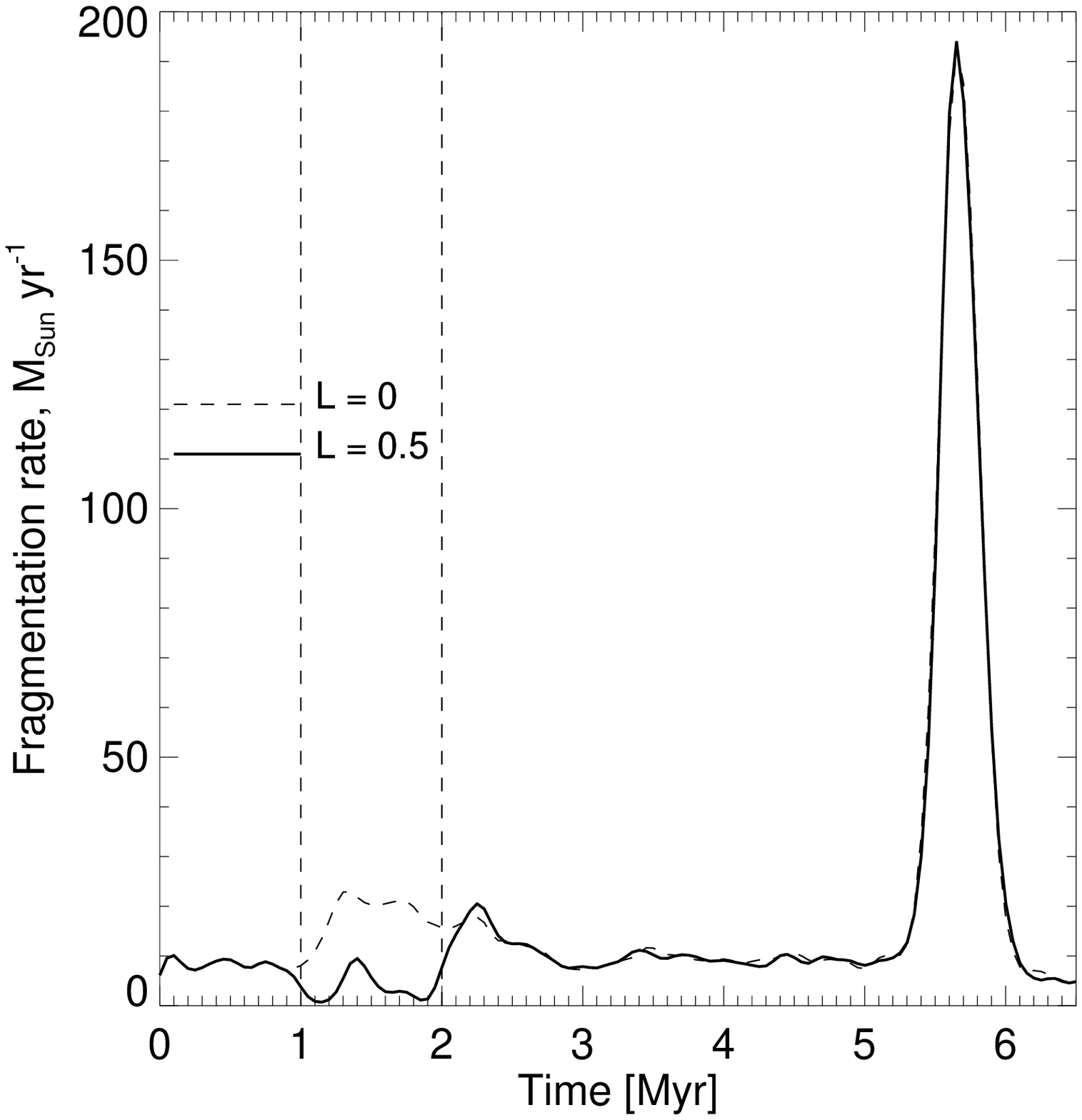}
    \includegraphics[trim = 6mm 27mm 6mm 9mm, clip, width=0.3 \textwidth]{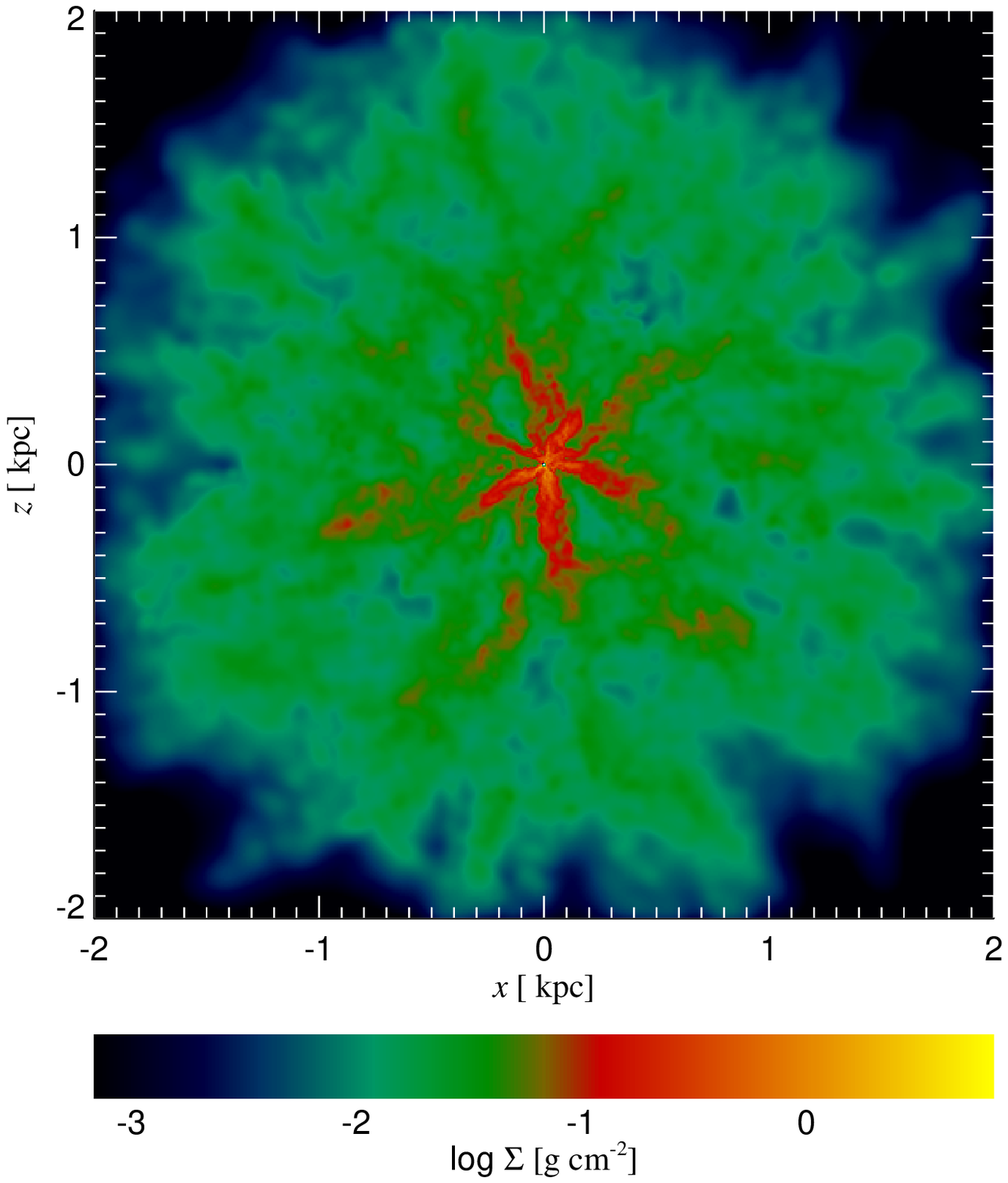}
    \includegraphics[trim = 6mm 27mm 6mm 9mm, clip, width=0.3 \textwidth]{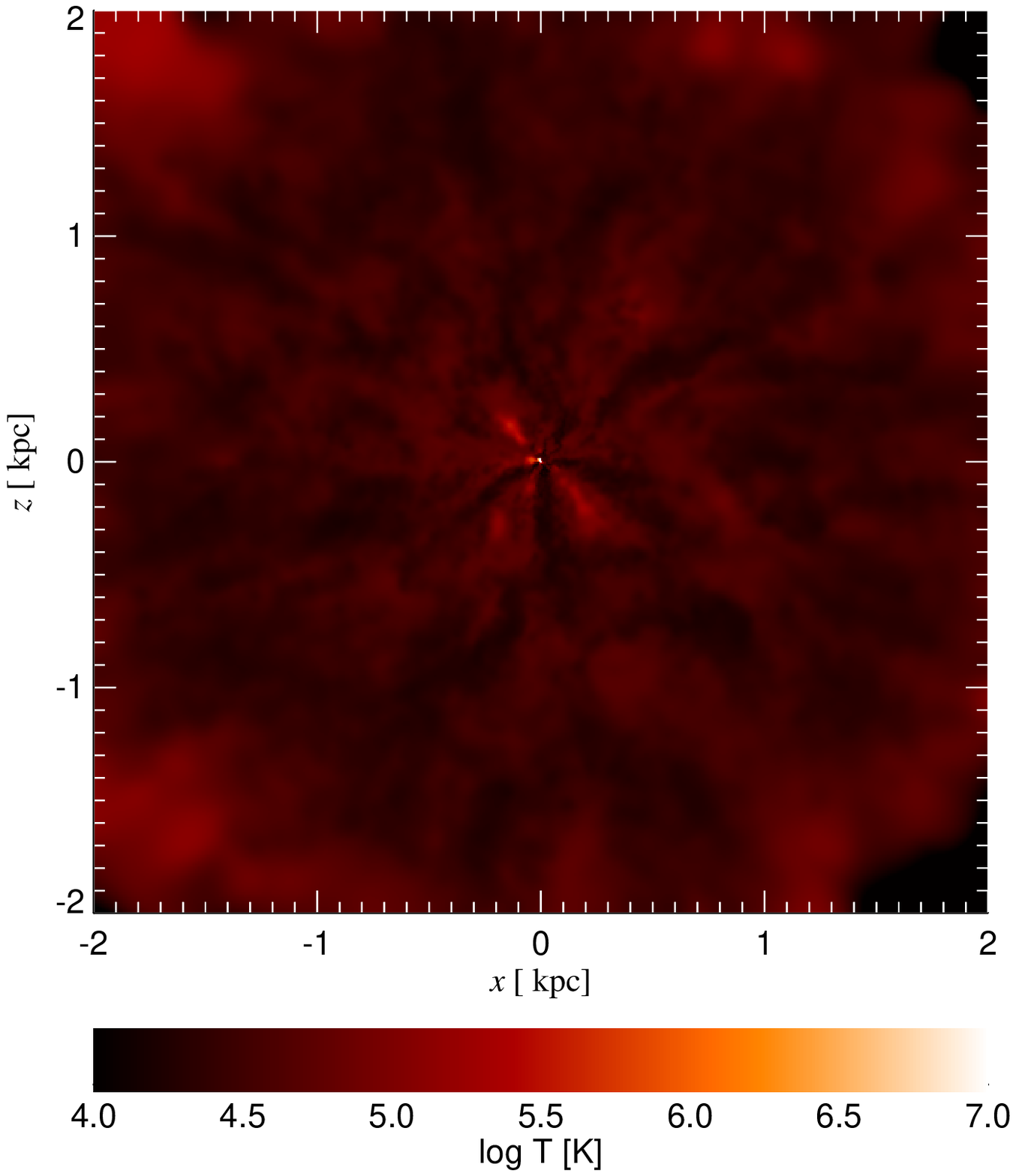}
    \includegraphics[trim = 0 0 0 0, clip, width=0.33 \textwidth]{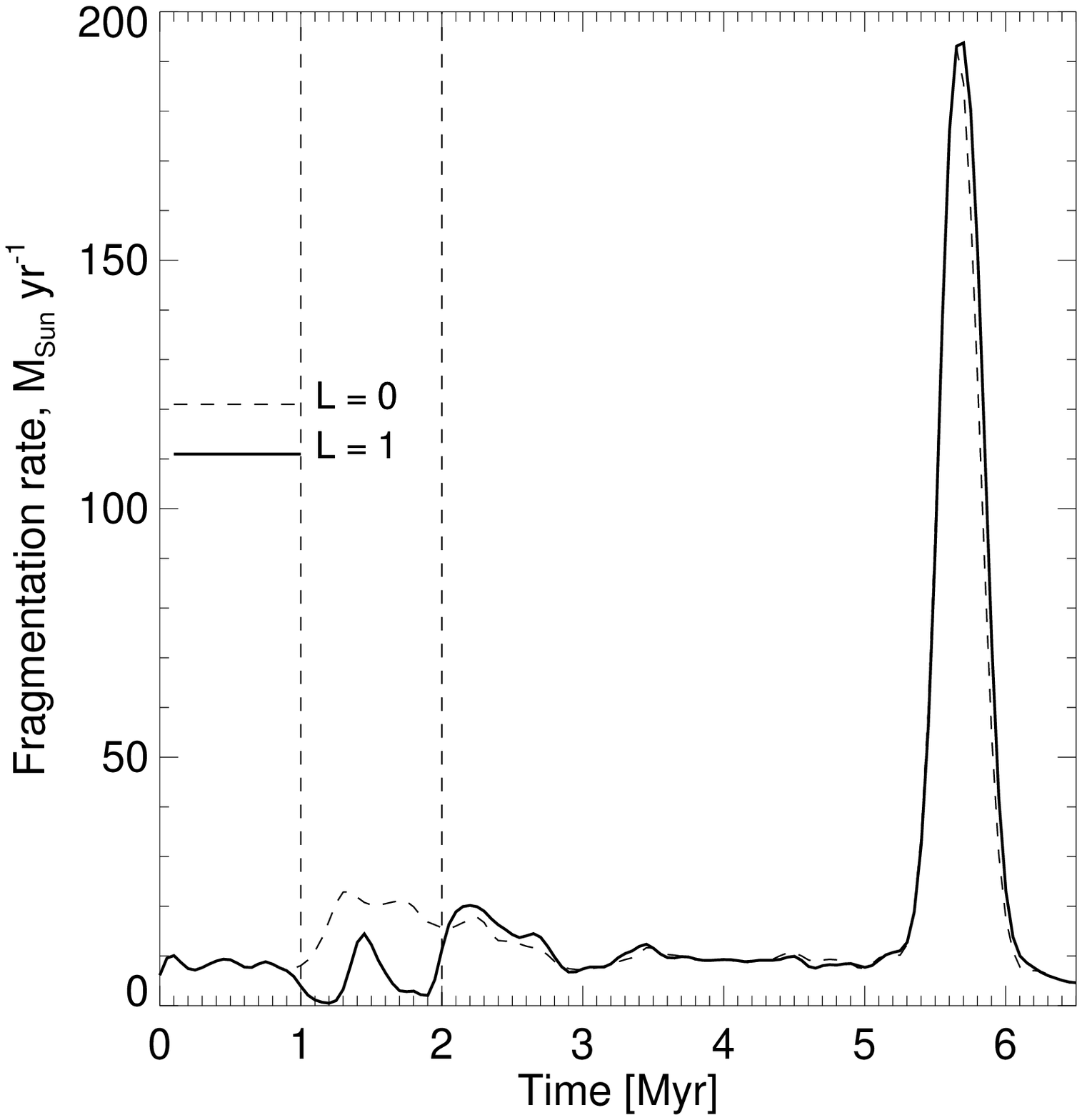}
    \includegraphics[trim = 6mm 27mm 6mm 9mm, clip, width=0.3 \textwidth]{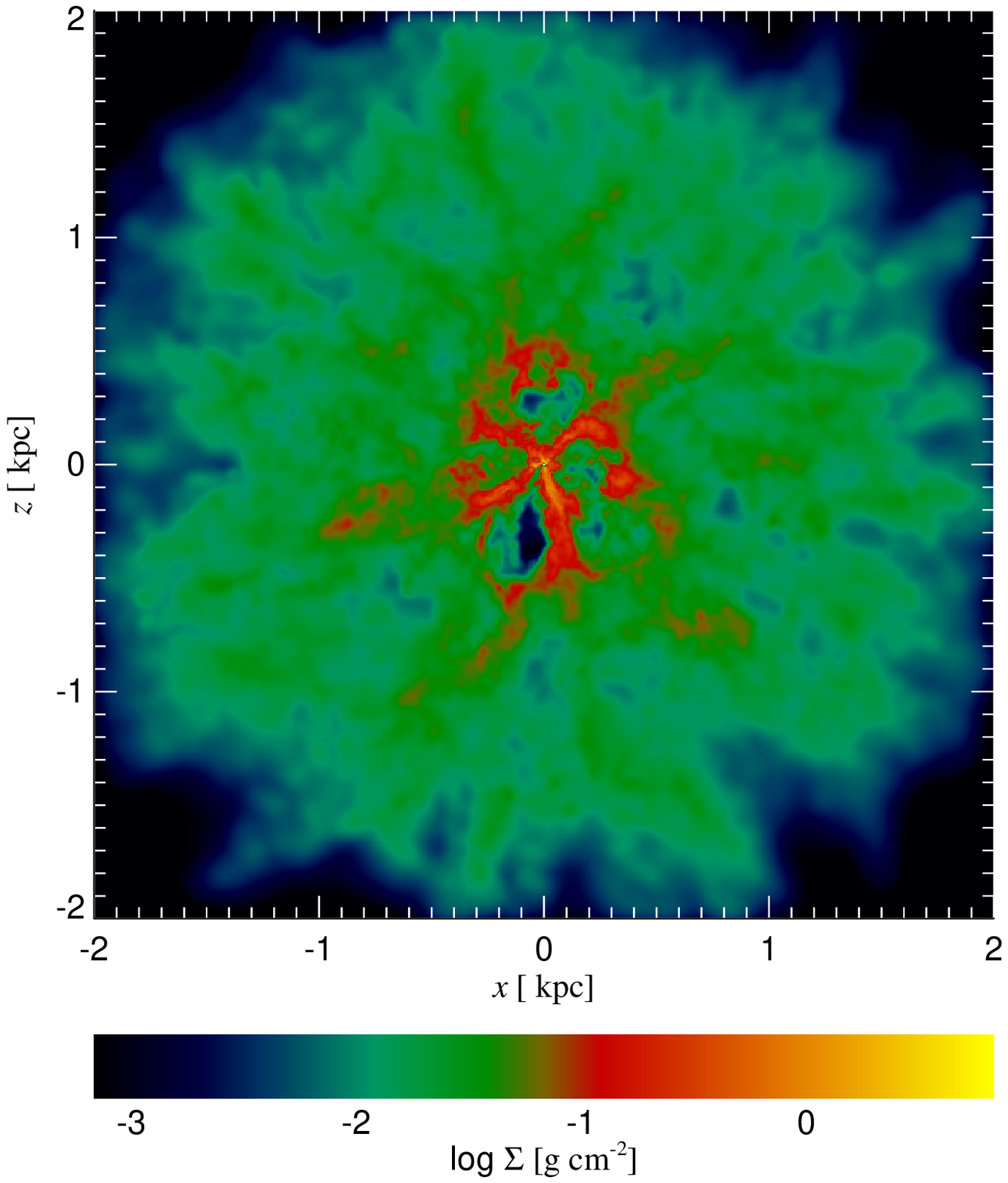}
    \includegraphics[trim = 6mm 27mm 6mm 9mm, clip, width=0.3 \textwidth]{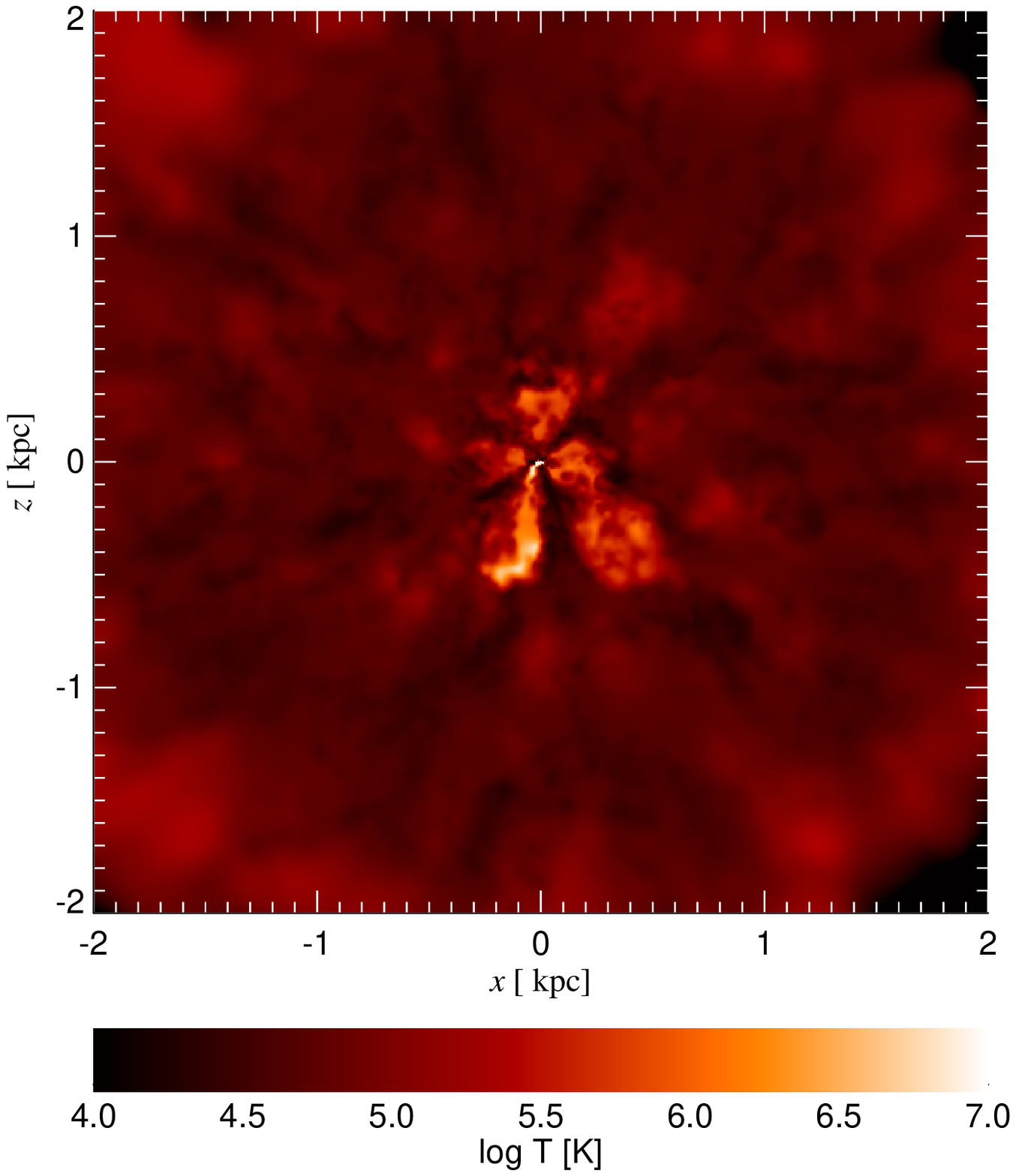}
    \includegraphics[trim = 0 0 0 0, clip, width=0.33 \textwidth]{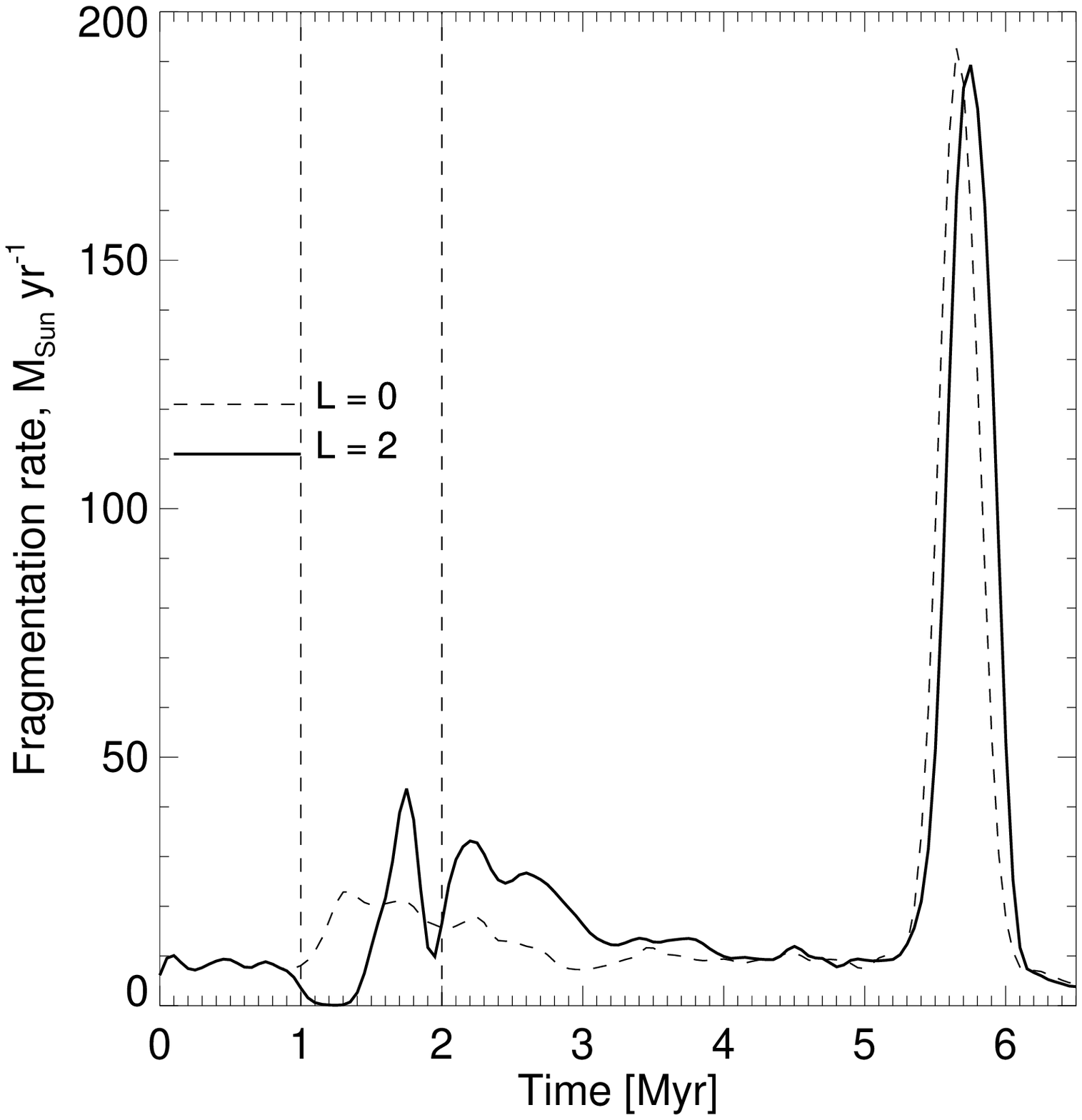}
  \caption{Density maps (left column), temperature maps (middle
    column), both at $t = 1$~Myr after the start of the AGN phase, and
    fragmentation rates over time in three of the seven simulations
    with $t_{\rm q} = 1$~Myr. Fragmentation rate plots include the
    rates of a control simulation (dashed lines). Vertical dashed
    lines in the fragmentation rate plots show the start and end of
    the AGN phase. From top to bottom, the simulations have
    progressively higher values of AGN luminosity $L = L_{\rm AGN} /
    \left(1.3\times 10^{46}{\rm erg s}^{-1}\right)$: $L = 0.5, 1,
    2$. See below for simulations with higher luminosities. }
  \label{fig:fragrates_T1}
\end{figure*}

\begin{figure*}
    \includegraphics[trim = 6mm 27mm 6mm 9mm, clip, width=0.3 \textwidth]{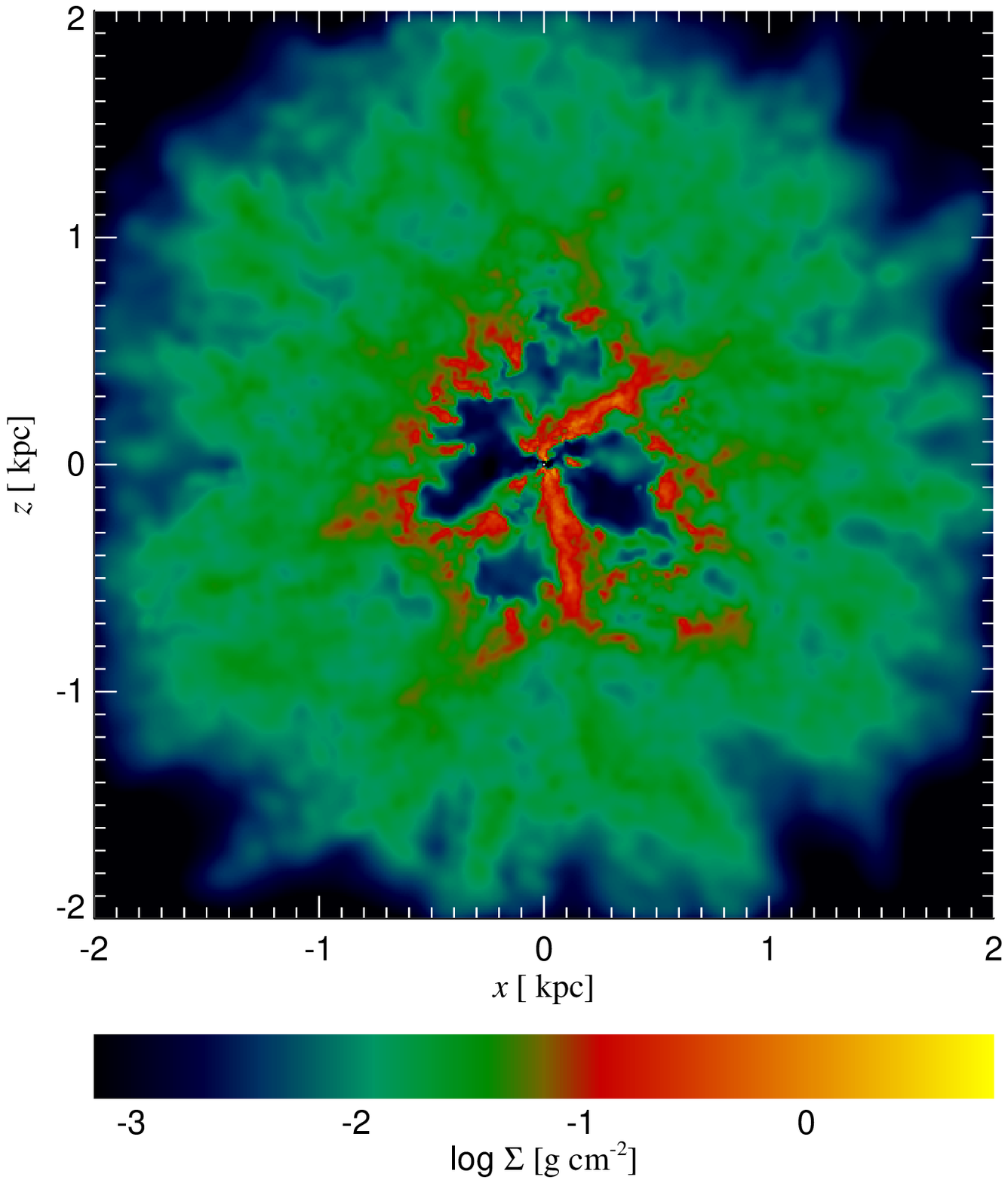}
    \includegraphics[trim = 6mm 27mm 6mm 9mm, clip, width=0.3 \textwidth]{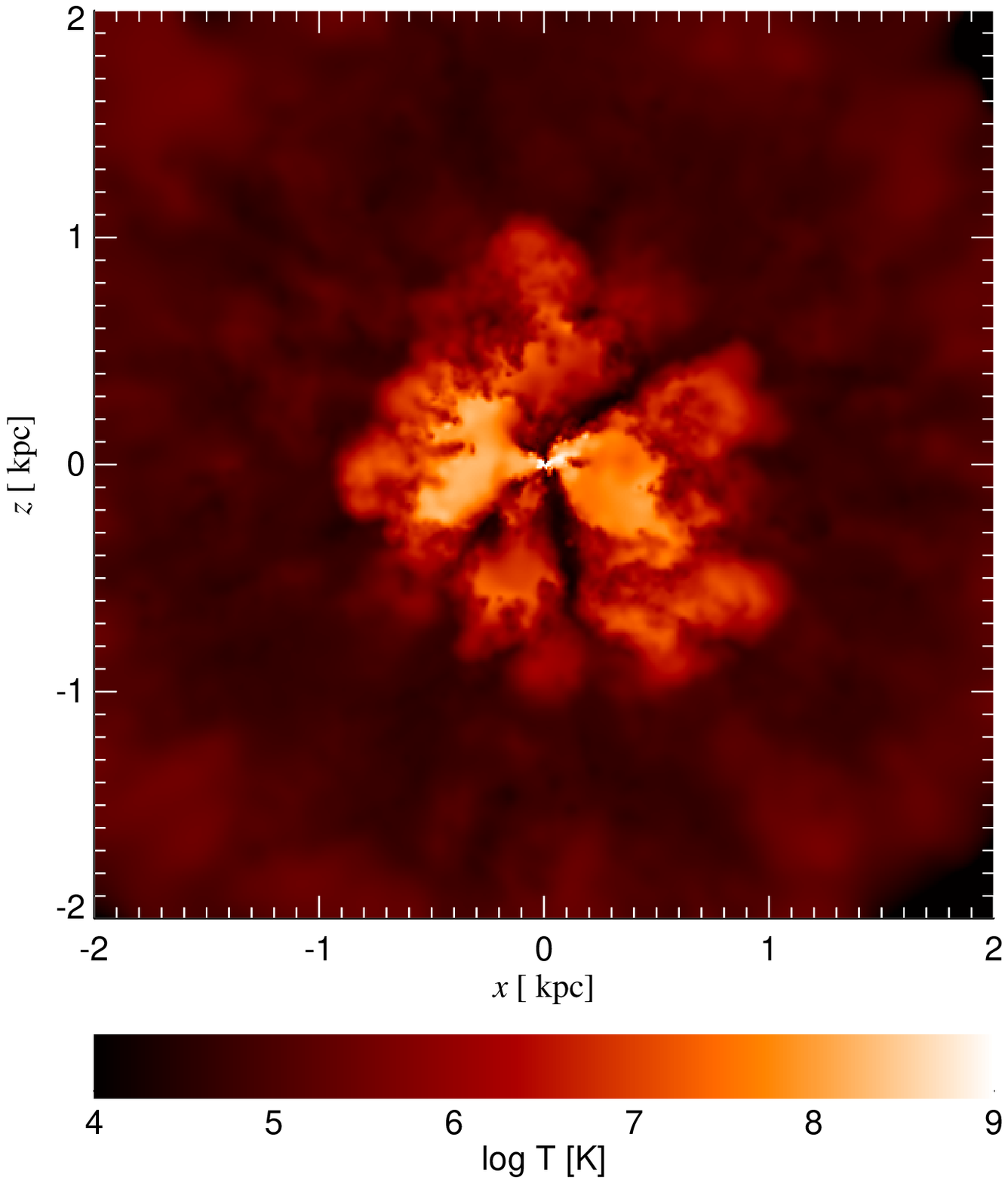}
    \includegraphics[trim = 0 0 0 0, clip, width=0.33 \textwidth]{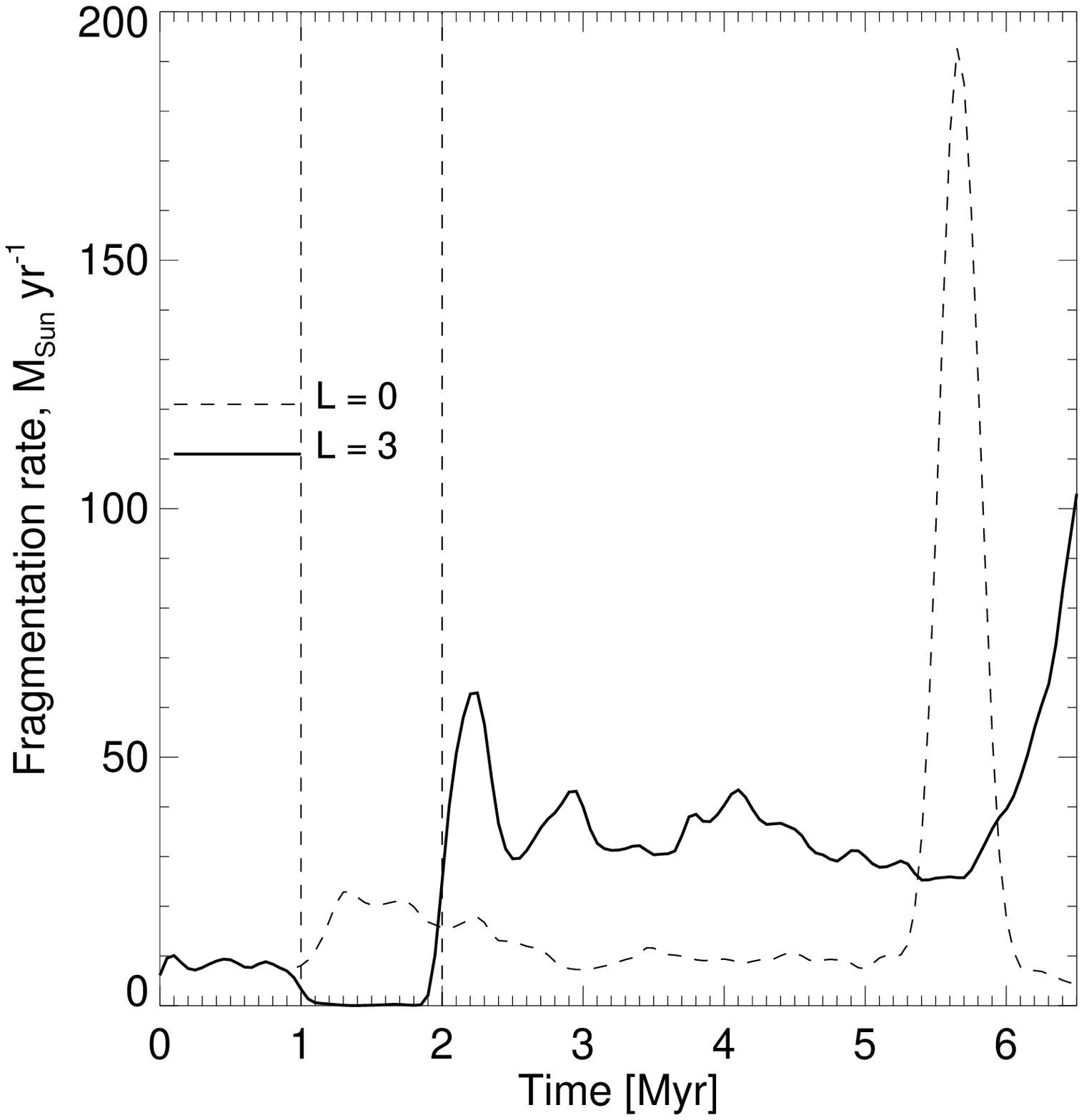}
    \includegraphics[trim = 6mm 27mm 6mm 9mm, clip, width=0.3 \textwidth]{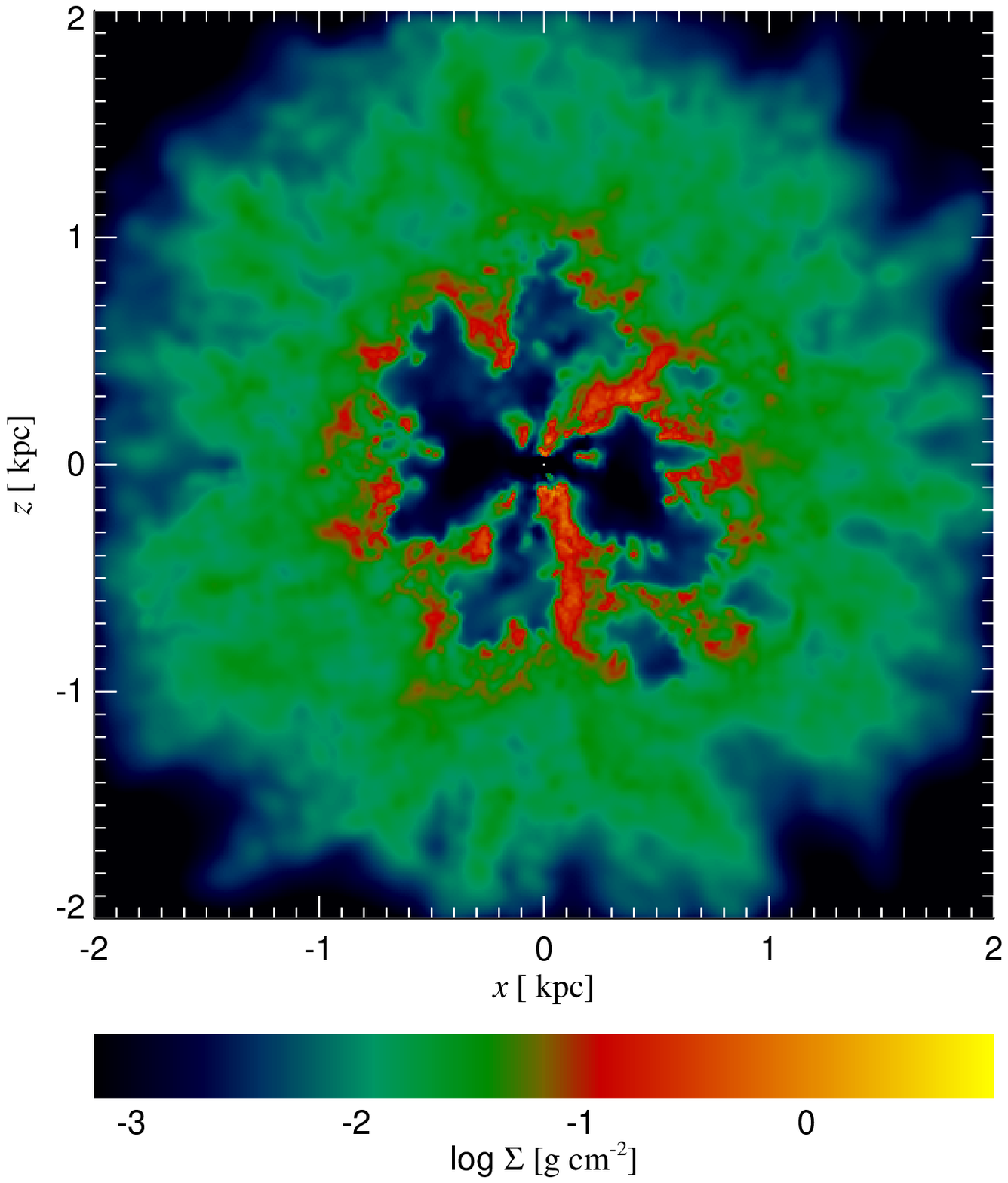}
    \includegraphics[trim = 6mm 27mm 6mm 9mm, clip, width=0.3 \textwidth]{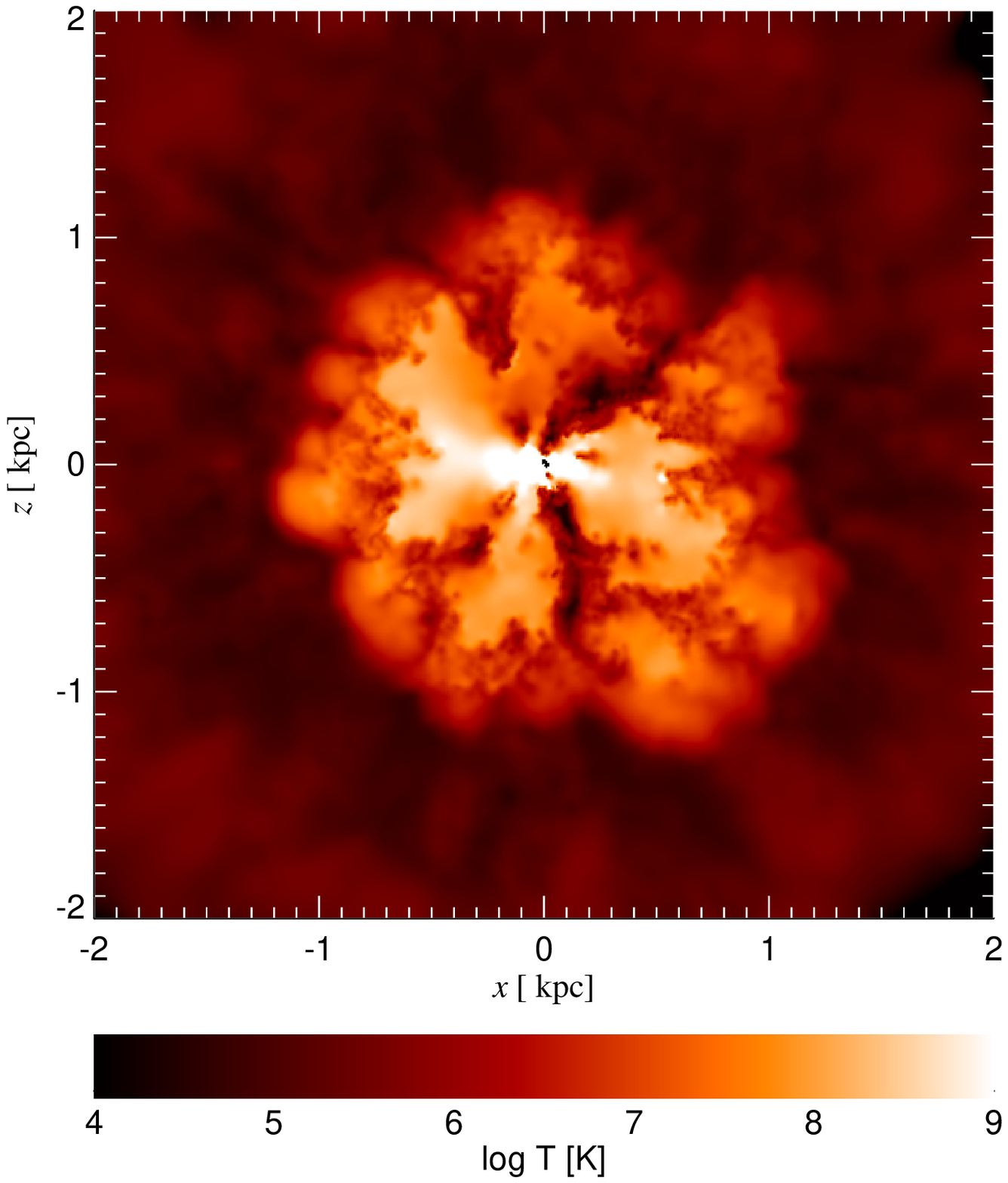}
    \includegraphics[trim = 0 0 0 0, clip, width=0.33 \textwidth]{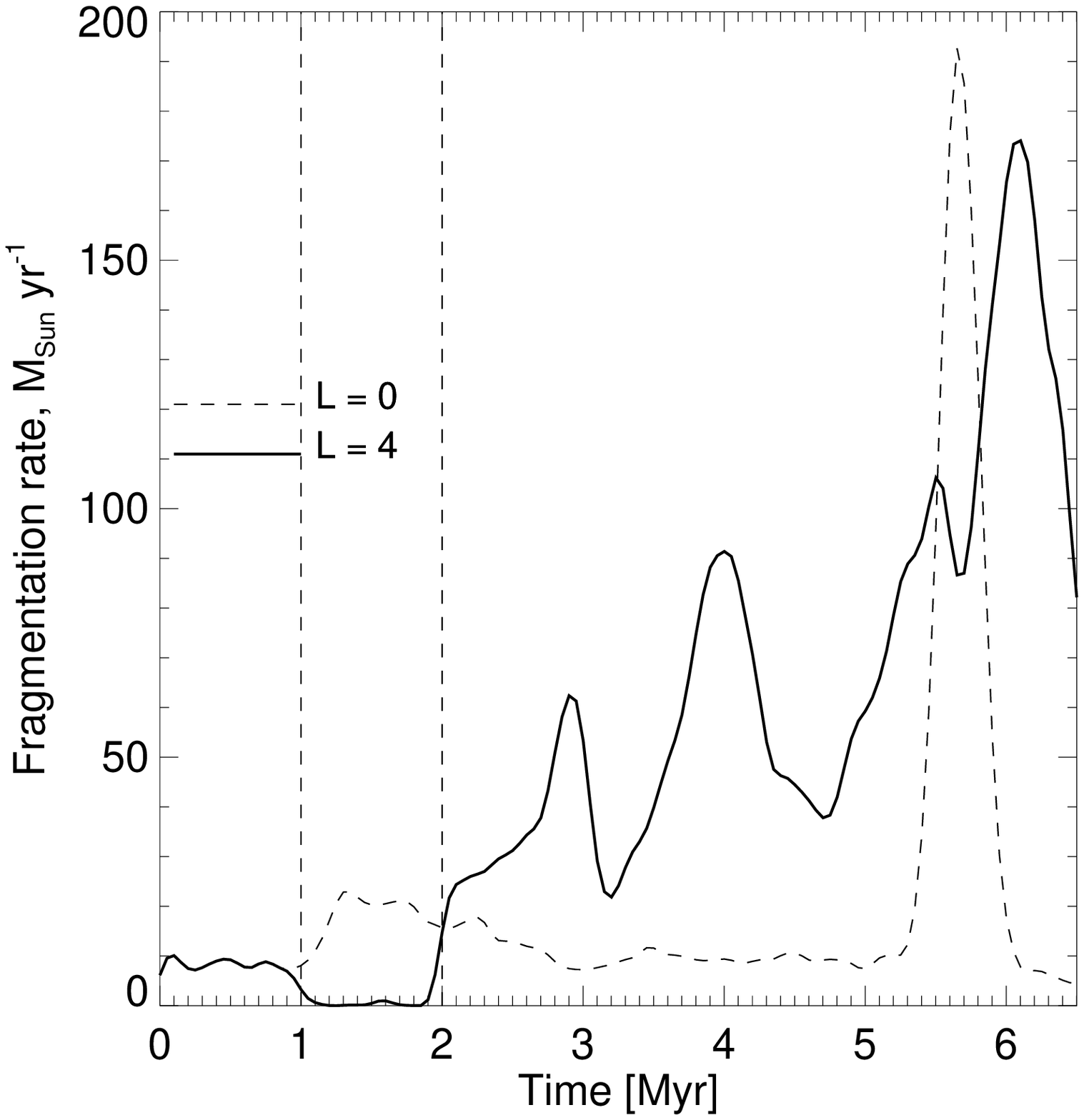}
    \includegraphics[trim = 6mm 27mm 6mm 9mm, clip, width=0.3 \textwidth]{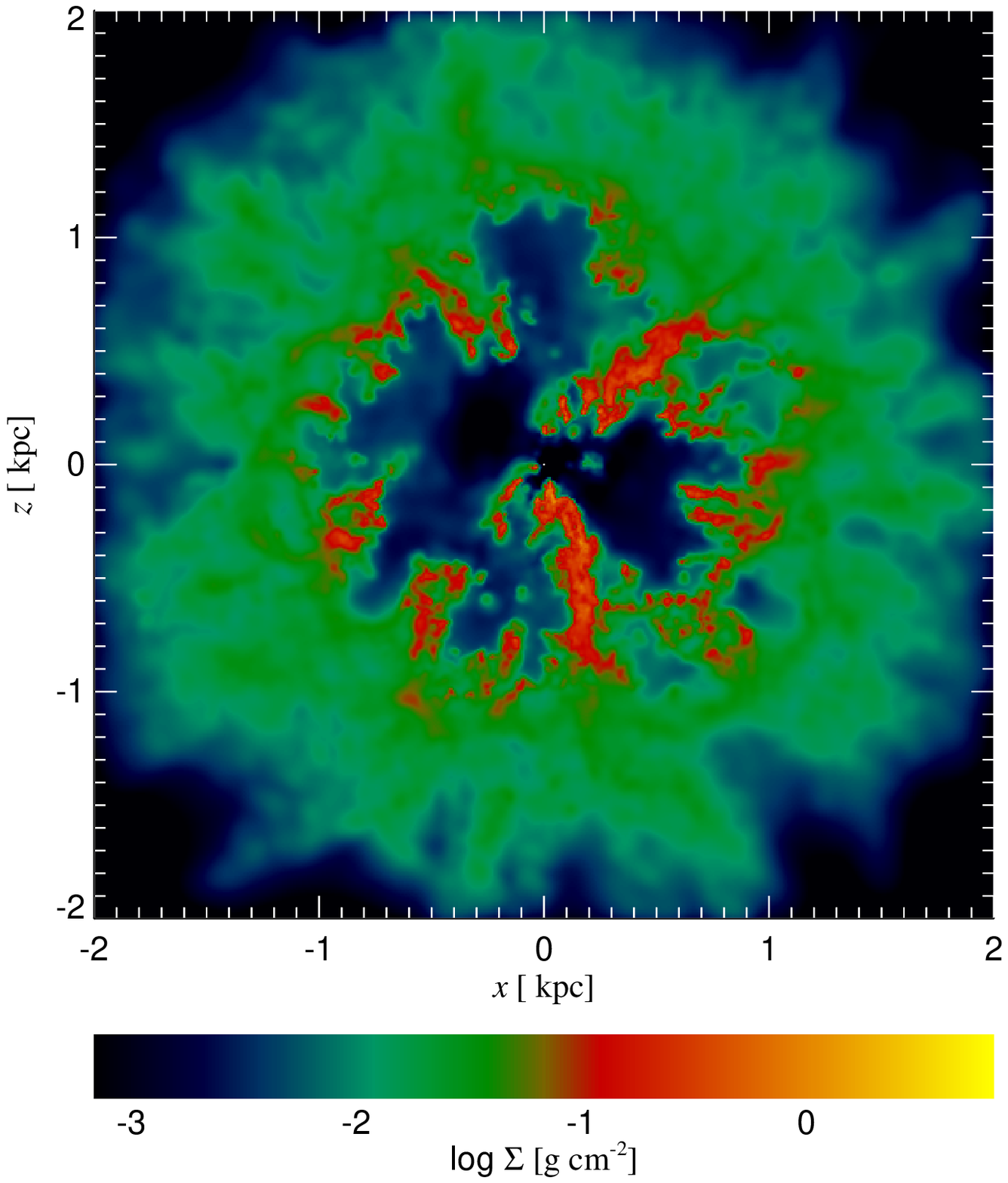}
    \includegraphics[trim = 6mm 27mm 6mm 9mm, clip, width=0.3 \textwidth]{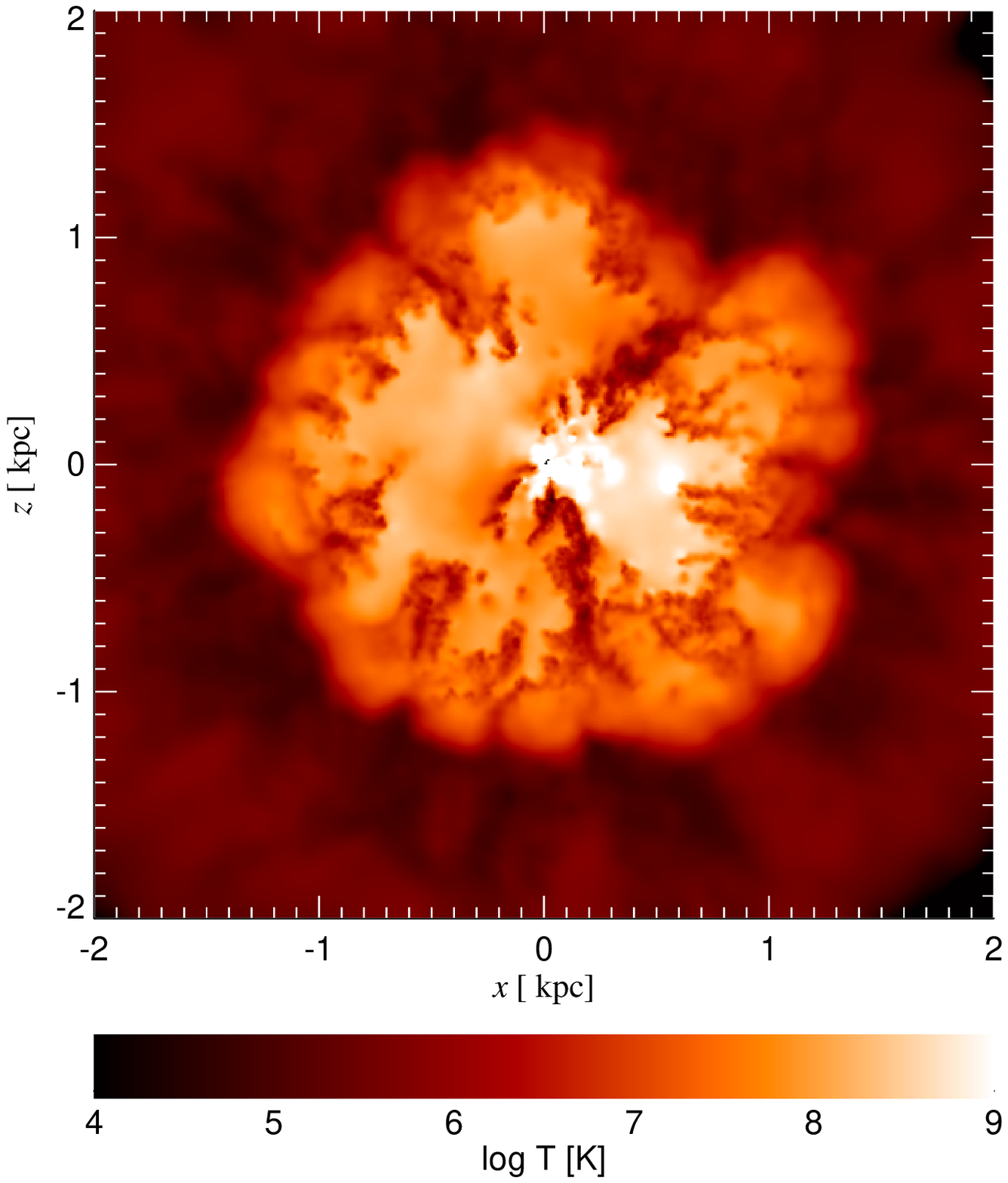}
    \includegraphics[trim = 0 0 0 0, clip, width=0.33 \textwidth]{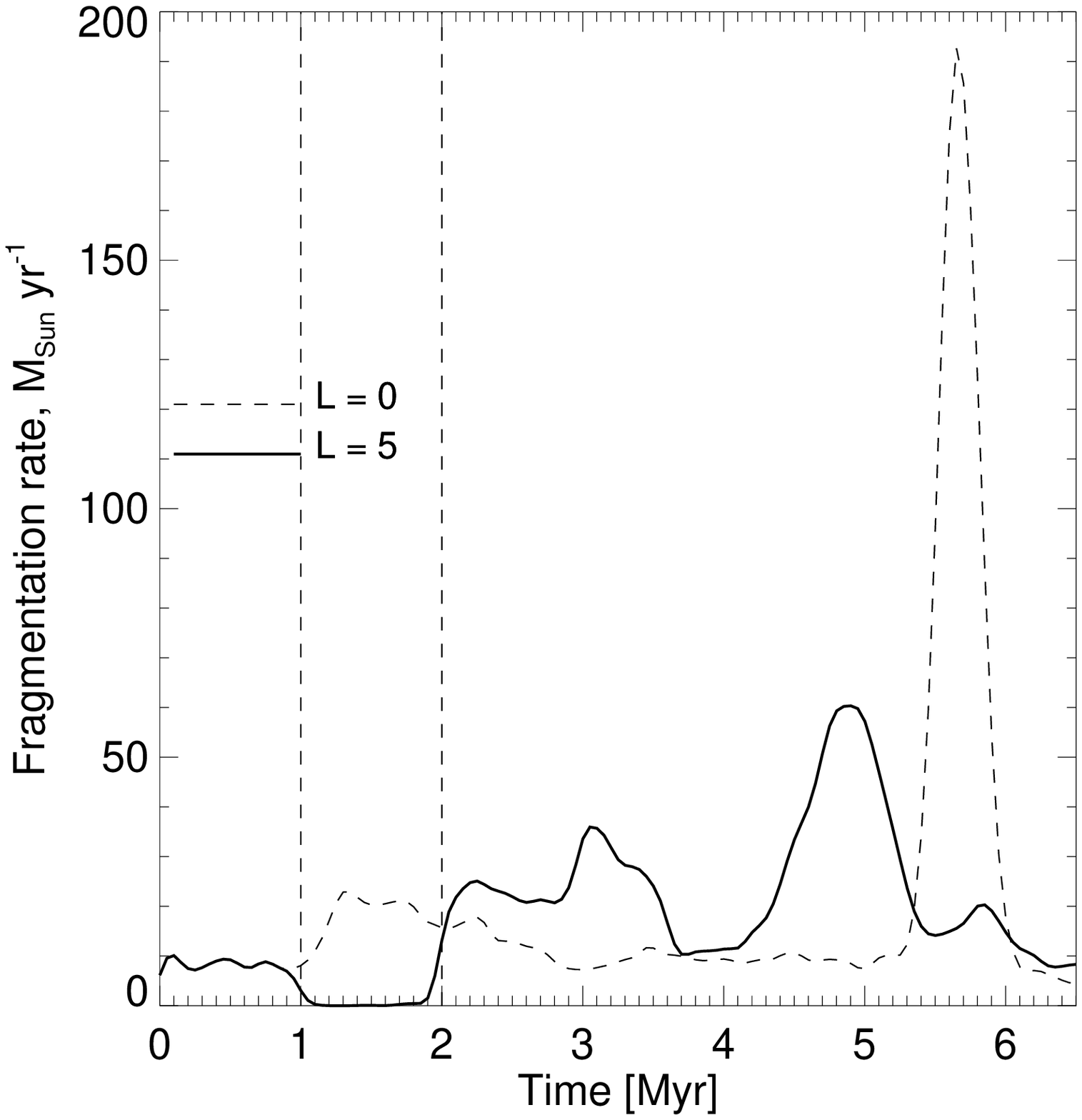}
    \includegraphics[trim = 6mm 27mm 6mm 9mm, clip, width=0.3 \textwidth]{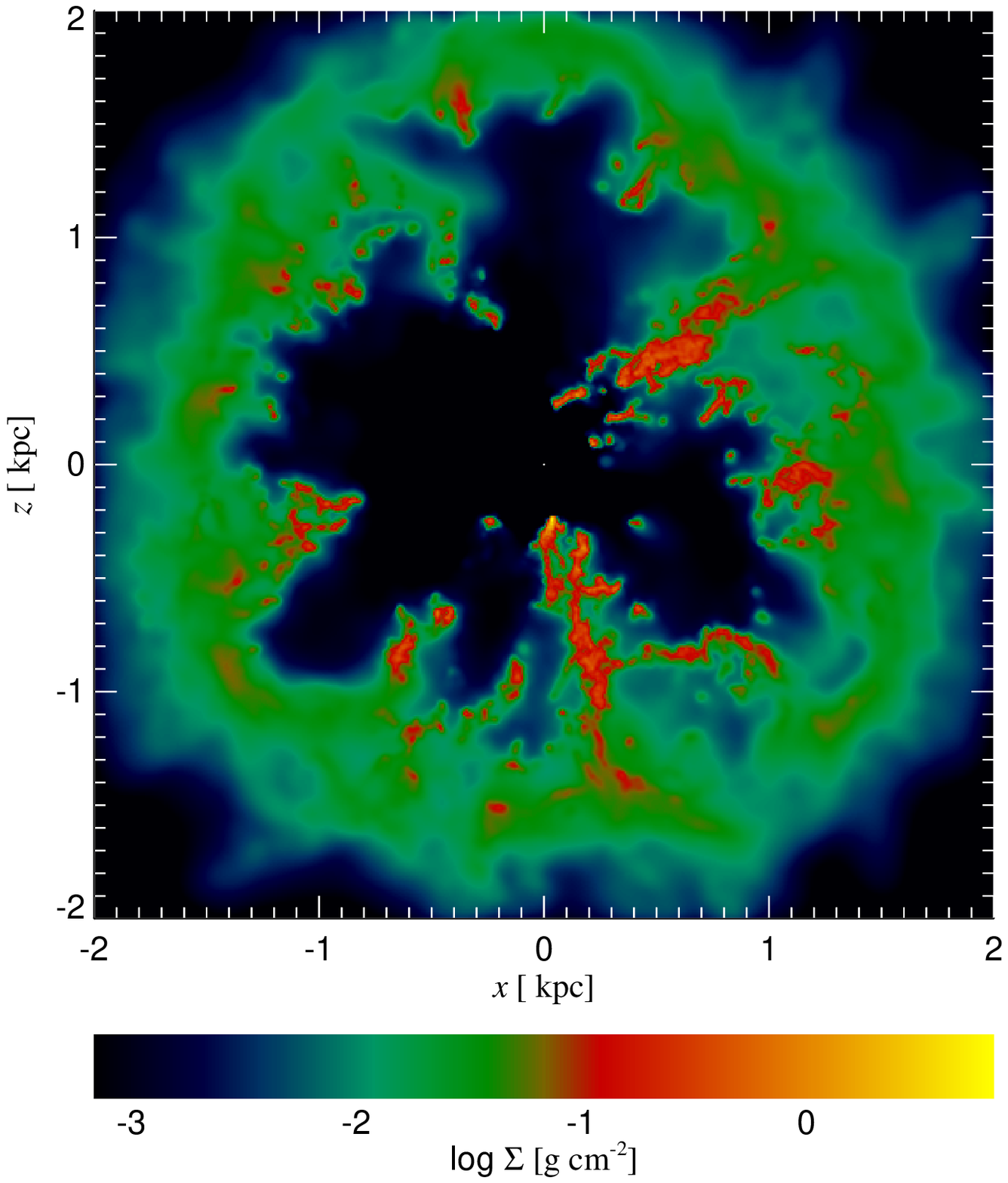}
    \includegraphics[trim = 6mm 27mm 6mm 9mm, clip, width=0.3 \textwidth]{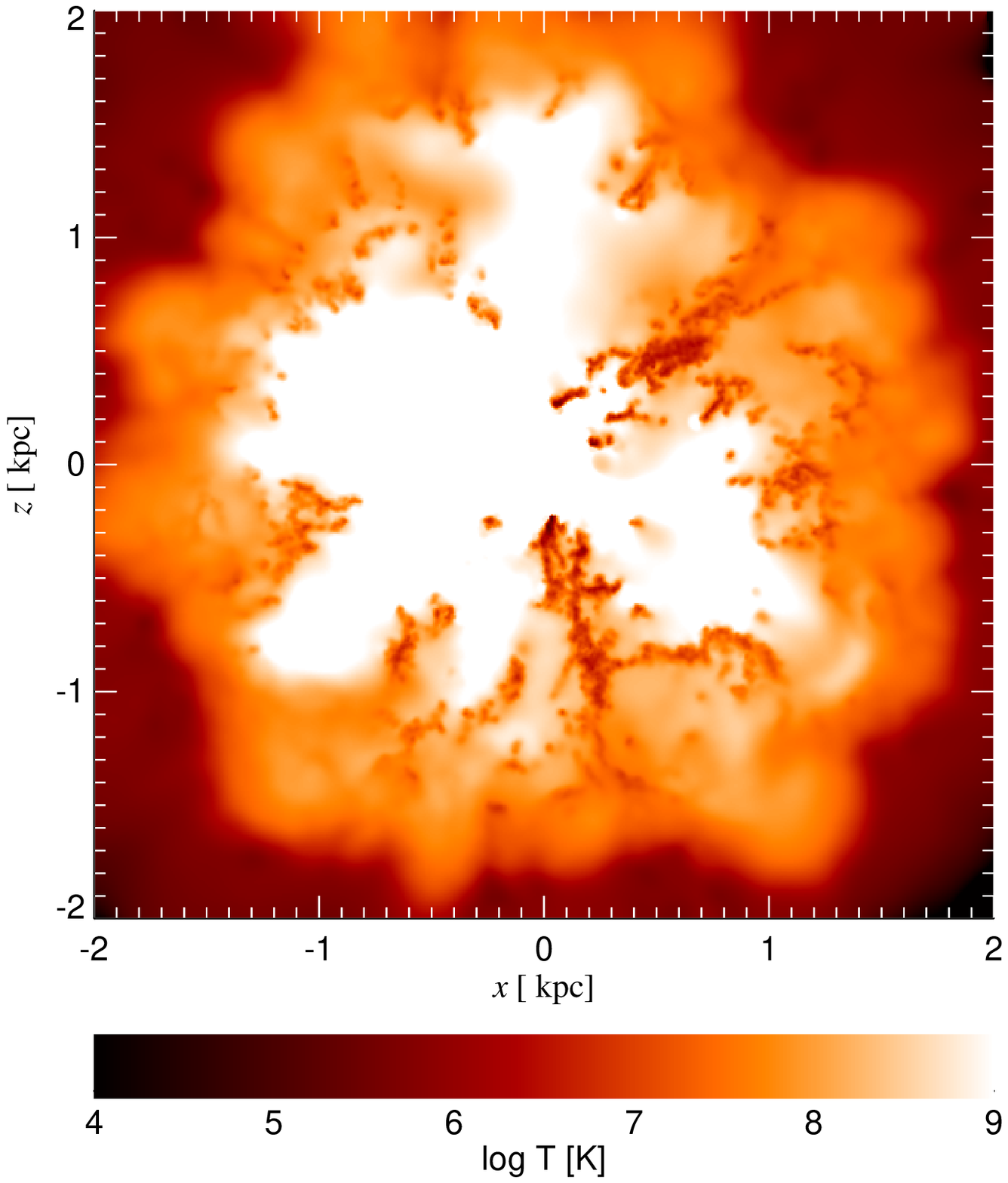}
    \includegraphics[trim = 0 0 0 0, clip, width=0.33 \textwidth]{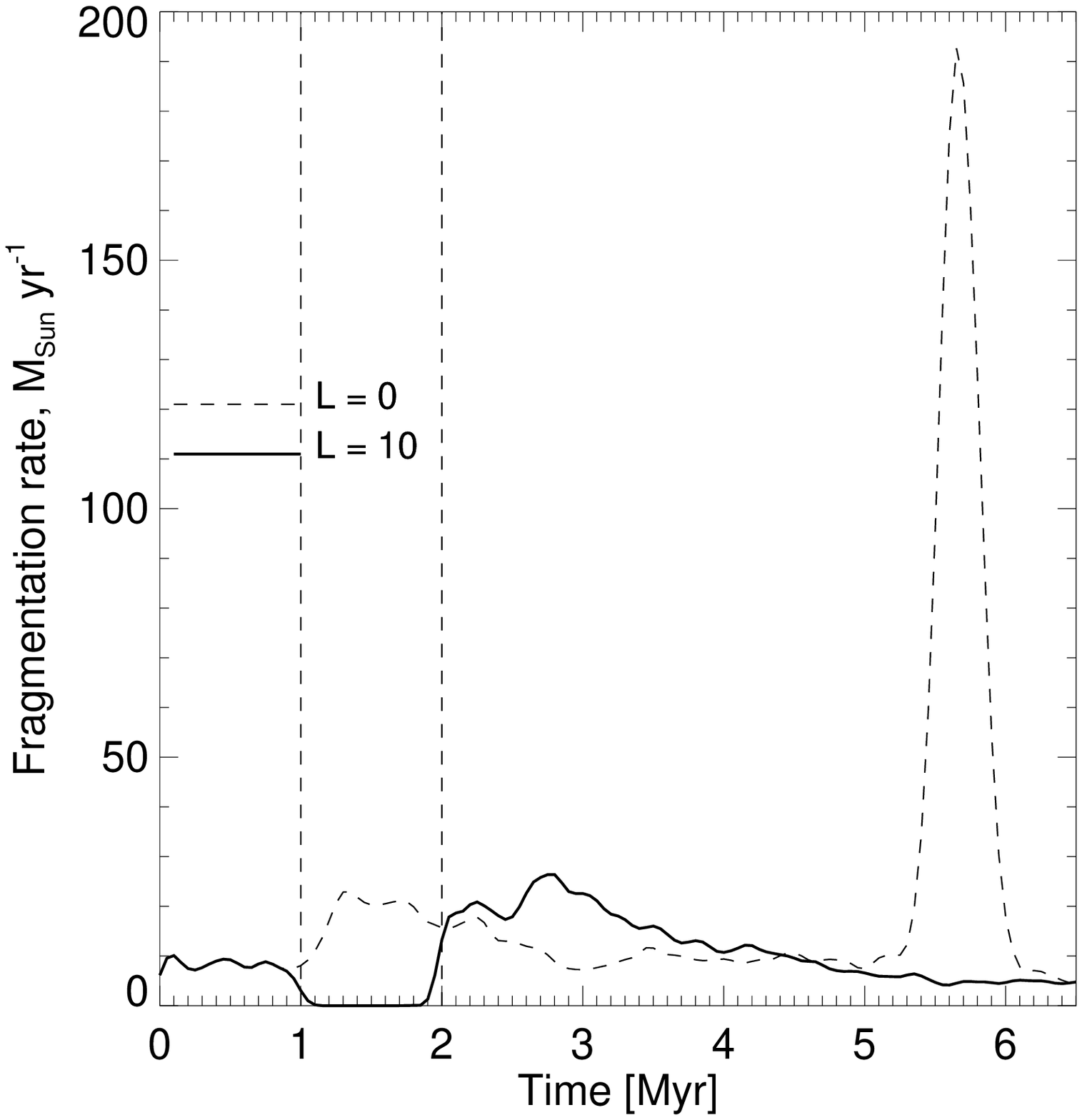}
  \caption{Continuation of Figure \ref{fig:fragrates_T1} for $L = 3,
    4, 5, 10$ from top to bottom. Note the different range of
    temperatures plotted.}
\end{figure*}

\begin{figure*}
  \centering
    \includegraphics[trim = 0 0 0 0, clip, width=1. \textwidth]{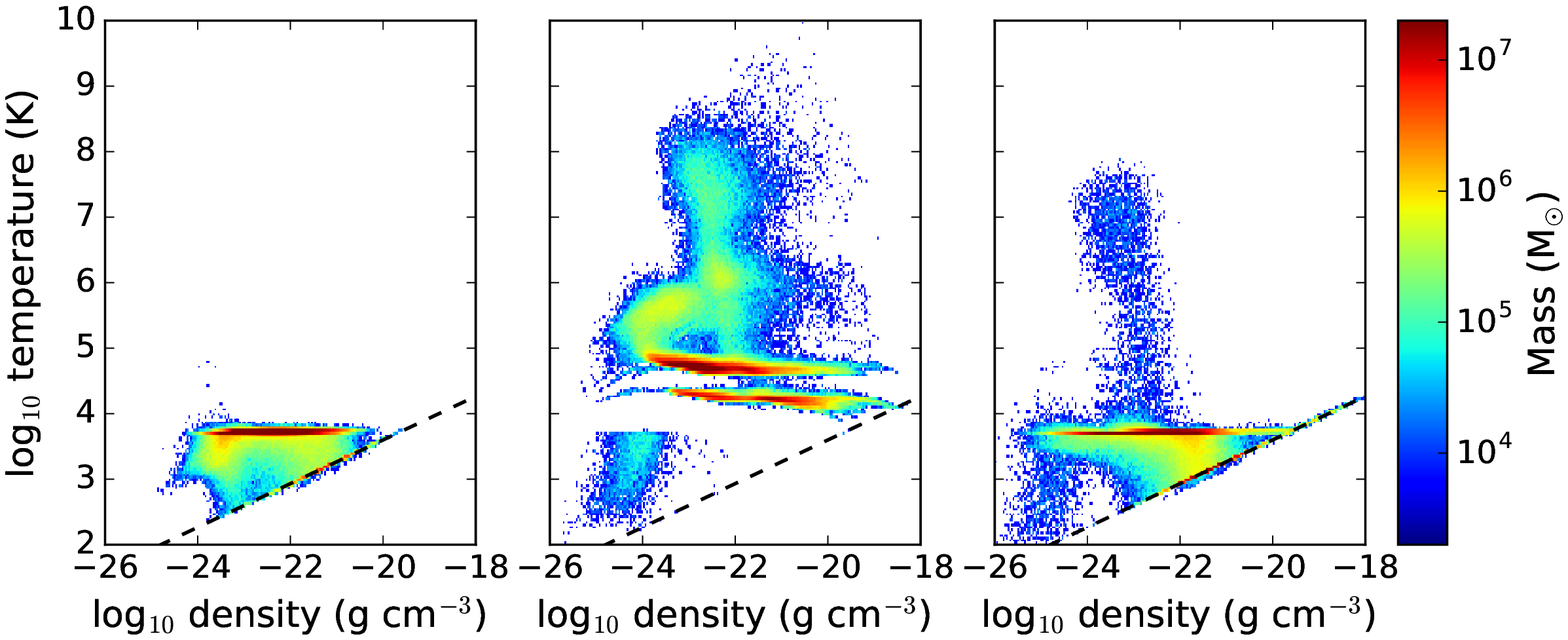}
  \caption{Density-temperature plots of gas in simulation L3T1, with
    $L = 3 \times 1.3 \times 10^{46}$~erg~s$^{-1}$ and $t_{\rm q} =
    1$~Myr at three different times: $t = 0.75$~Myr on the left
    (before the AGN switches on), $t = 1.75$~Myr in the middle (close
    to the end of the AGN phase), $t = 2.75$~Myr on the right (after
    the AGN has switched off). Colours indicate relative density of
    particles, with red being highest.}
  \label{fig:phaseplots}
\end{figure*}

In all simulations, the gas is allowed to relax for $1$~Myr before the
AGN is switched on. This creates a turbulent density structure, with
the ratio between the highest and lowest effective gas fraction
($f_{\rm g,eff} \equiv \rho_{\rm g}/\left(\rho_{\rm g} + \rho_{\rm
  pot}\right)$) of order 30. Approximately $7.6\%$ of the gas is
swallowed by the SMBH sink particle and $0.15\%$ of the gas is turned
into sink particles during this time.

Once the AGN switches on, an outflow begins to propagate through the
system, provided that the AGN luminosity is large enough (see Figure
\ref{fig:fragrates_T1}, left panels). It is unable to stop the
accretion of the densest gas, and produces bubbles which contain most
of the input energy. Dense filaments form in between the bubbles and
continue feeding the SMBH. As expected, the bubbles are more
pronounced and the mass outflow rates are higher in the higher
luminosity simulations.

The density contrast increases during AGN activity, as material is
removed from some parts of the simulation but not others. In the L0.5,
L1 and L2 simulation pairs (all pairs evolve identically until $t =
2$~Myr), the density contrast is not much higher at $t = 2$~Myr than
in the control simulation at the same time. In L3 simulations, the
ratio between highest and lowest effective gas fractions increases to
$\sim 300$, i.e. by a factor ten, from $t = 1$~Myr to $t =
2$~Myr. This ratio is $\sim 1000$ in L4 simulations, $\sim 2000$ in L5
and $\sim 10^4$ in L10. The highest effective gas fraction is always
at the radius where the outer edges of the outflow bubbles are. As
expected, the more powerful the AGN is, the more effective it is at
compressing gas at the leading edge of the outflow bubble. If the AGN
was active for a longer time, or if the mean gas density was lower,
these bubbles would escape from the gas shell and the density contrast
would decrease.

\begin{figure}
  \centering
    \includegraphics[trim = 0 0 0 0, clip, width=0.49 \textwidth]{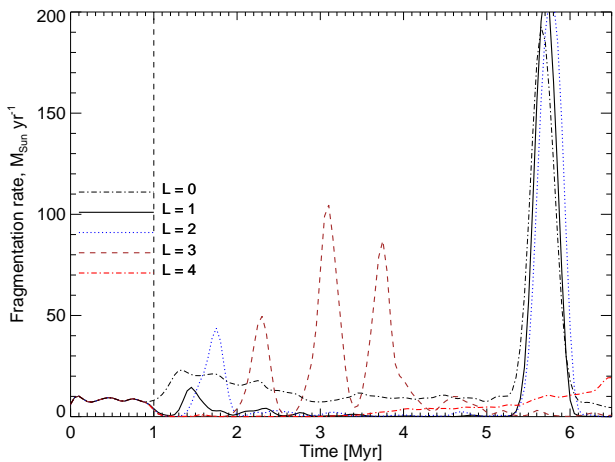}
  \caption{Fragmentation rates in five simulations with $t_{\rm q} =
    \infty$ (thick lines) plus a control simulation (thin dashed
    line). Vertical dashed line shows the beginning of the AGN
    phase. The lines are labelled with values of $L = L_{\rm AGN} /
    \left(1.3\times 10^{46}{\rm erg s}^{-1}\right)$: $L = 0, 1, 2, 3,
    4$. As expected, high-luminosity AGN stop all fragmentation, while
    low-luminosity ones have only a marginal effect.}
  \label{fig:fragrates_inf}
\end{figure}

\subsection{Integrated fragmentation rates}

The right panels of figure \ref{fig:fragrates_T1} show the
fragmentation rates in the simulations with AGN phase duration of
$1$~Myr, plus a control simulation with no AGN. All fragmentation
rates in this and subsequent figures are sampled every $5\times
10^4$~yr, but smoothed using a normalised kernel with weights
$\left\{1,4,6,4,1\right\}$ in order to reduce spurious noise. The
fragmentation rate before the AGN switches on is a few
$\msun$~yr$^{-1}$, and remains at that rate in the control simulation
until $t \simeq 1.1$~Myr and then increases to $\sim 20
\msun$~yr$^{-1}$, due to material falling inwards and reaching higher
densities. Later, as material is consumed, the fragmentation rate
drops down to its initial values, except for a significant peak at
$\sim5.7$~Myr, which happens because turbulent motions decay and are
no longer able to support the gas against dynamical collapse toward
the centre.

In the simulations with $L > 0$, AGN activity rapidly suppresses
fragmentation. In L3T1, L4T1 and L5T1 simulations, this suppression
lasts until the AGN switches off. The suppression is caused by two
effects: AGN heating evaporates some clumps and heats the gas overall,
slowing its fragmentation (see Figure \ref{fig:fragrates_T1}, middle
panels), while the AGN wind outflow pushes gas away from the centre,
thus diluting it. In the L0.5T1, L1T1 and L2T1 simulations, the wind
is not powerful enough to prevent gas accretion toward the centre, and
dense filaments of inflowing gas form and begin fragmenting even
before the AGN switches off. The fragmentation rate is higher in the
L2T1 simulation than in the L1T1 because the filaments are further
compressed by the outflow bubbles. When the densest gas fragments, the
less dense leftover material is again pushed away by the outflow and
further fragmentation decreases until the AGN switches off. At later
times, simulations L0.5T1, L1T1 and L2T1 follow closely the control
simulation, because the effect of the outflow upon the global gas
properties was negligible.

\begin{figure*}
  \centering
    \includegraphics[trim = 0 0 0 0, clip, width=0.49 \textwidth]{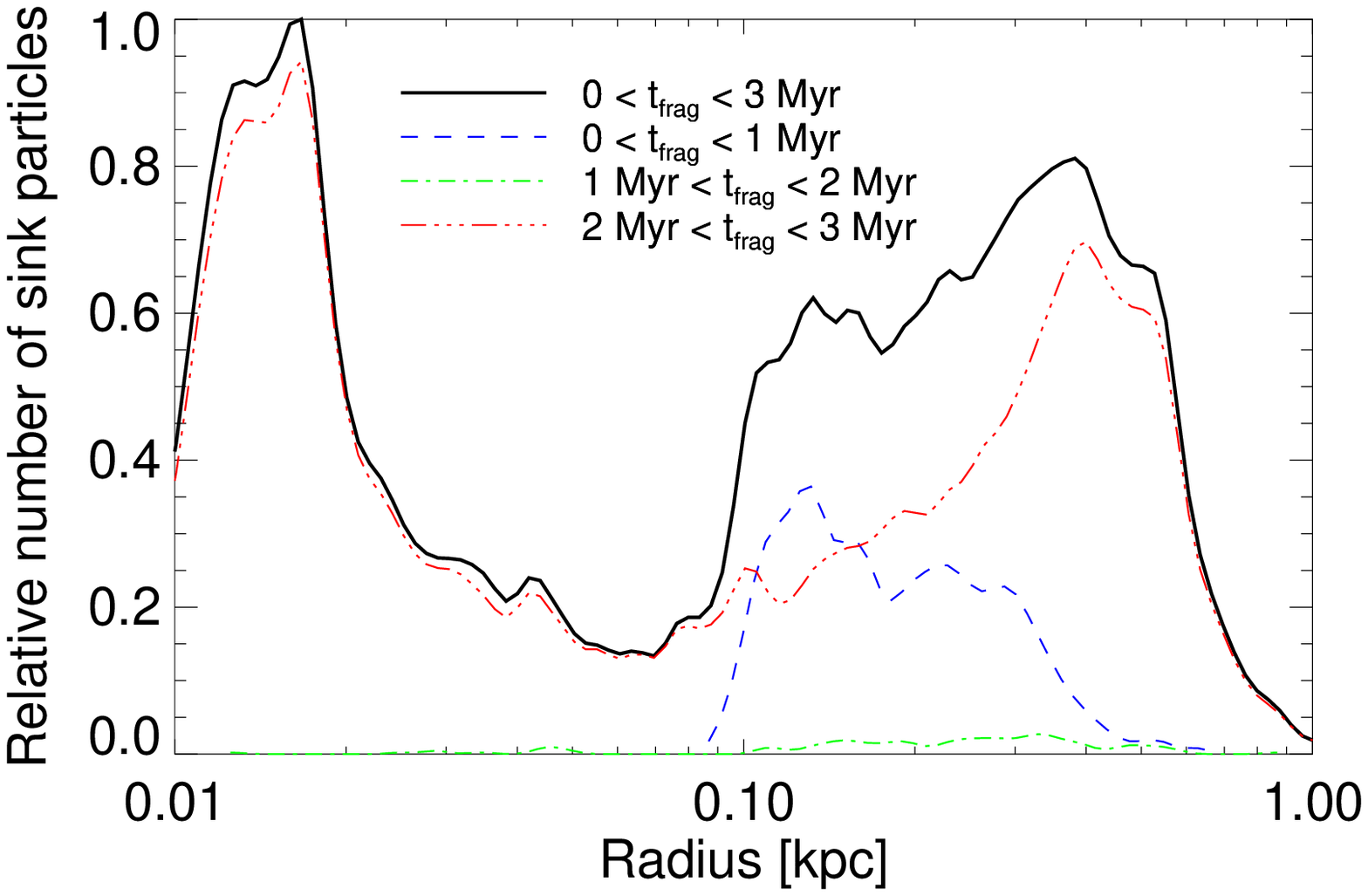}
    \includegraphics[trim = 0 0 0 0, clip, width=0.49 \textwidth]{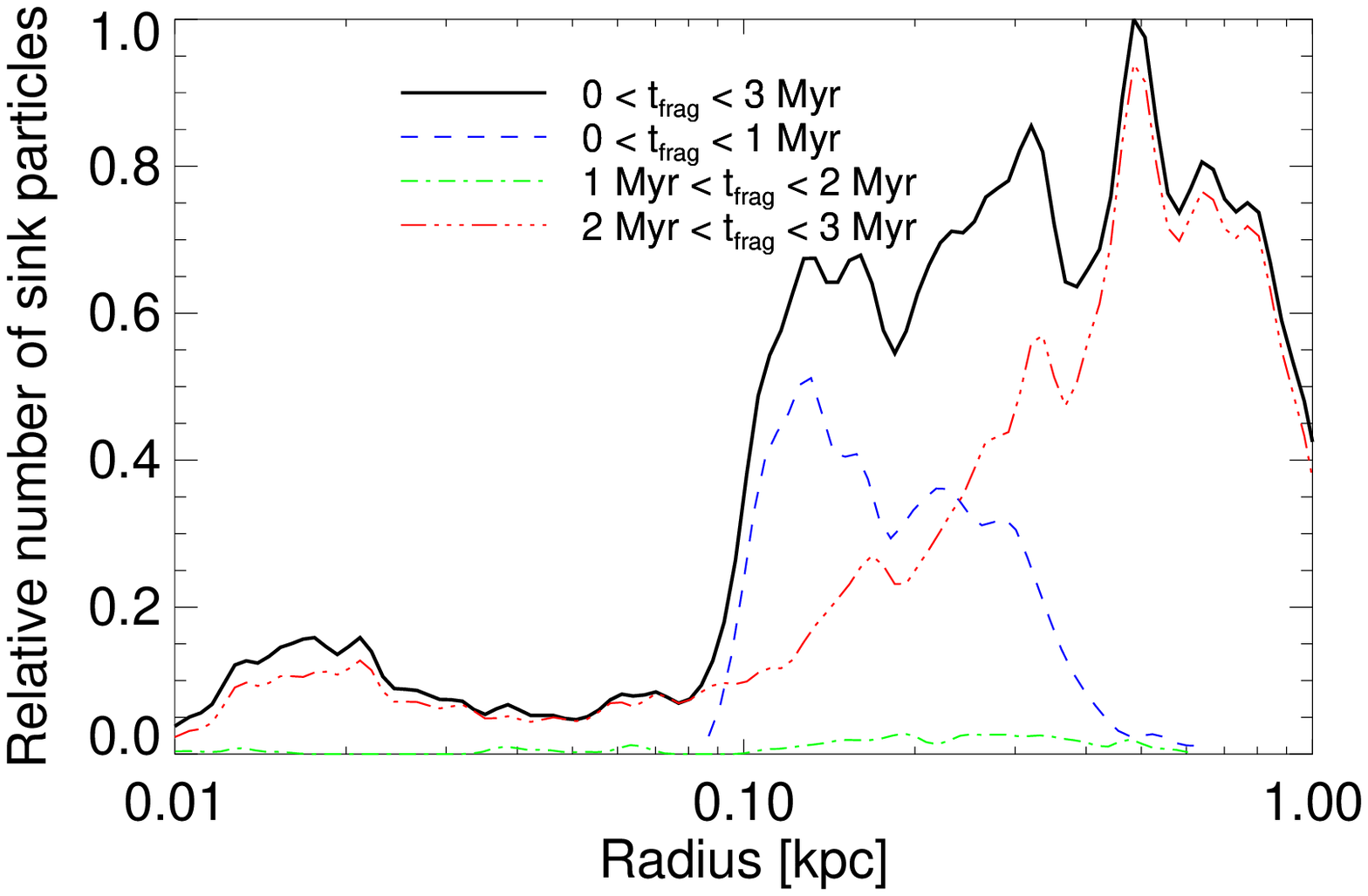}
  \caption{Distribution of the radial coordinates of sink particles at
    the time of their appearance (black solid lines), and its division
    into three time bins: particles that appear before the AGN
    switches on (blue dashed line), particles that appear during the
    AGN episode (green dot-dashed line) and particles that appear
    after the AGN switches off (red triple-dot-dashed line). Left
    panel shows simulation L3T1, right panel shows L5T1.}
  \label{fig:fragloc}
\end{figure*}

\begin{figure*}
  \centering
    \includegraphics[trim = 0 0 0 0, clip, width=0.49 \textwidth]{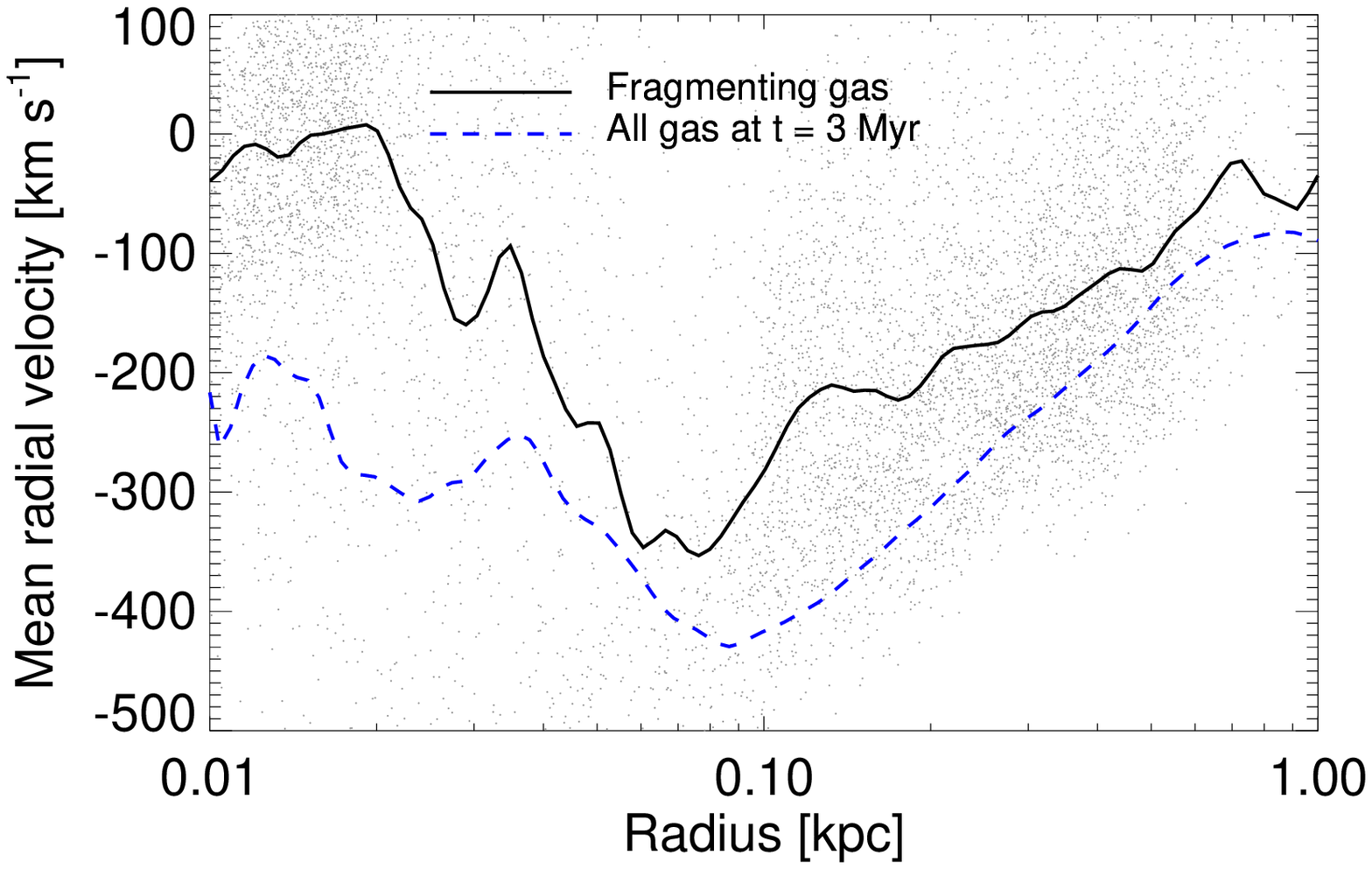}
    \includegraphics[trim = 0 0 0 0, clip, width=0.49 \textwidth]{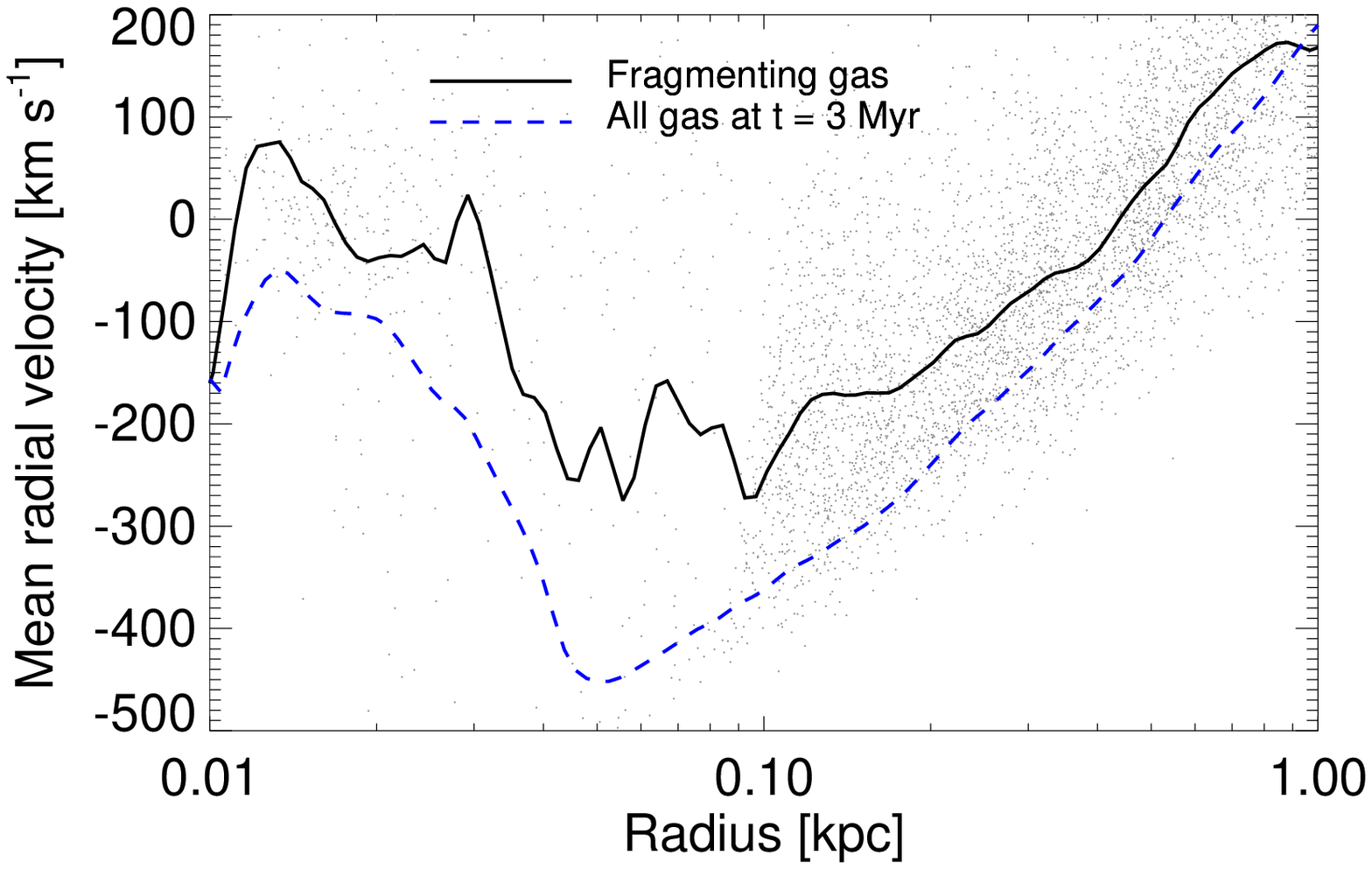}
  \caption{Mean radial velocity as a function of radius for sink
    particles at the time of their appearance (black solid lines),
    together with radial velocities of gas at $t = 3$~Myr (blue dashed
    line). Gray points are the radii and radial velocities of
    individual sink particles as they appear. Left panel shows
    simulation L3T1, right panel shows L5T1.}
  \label{fig:fragvel}
\end{figure*}

Once the AGN switches off, the outflow stalls and begins collapsing
back to the centre. In the L3T1 simulation, fragmentation resumes
immediately and its rate increases to $\sim 70 \msun$~yr$^{-1}$, as
the dense filaments that fragment had already begun their fall toward
the centre even before the AGN switched off (see the discussion
regarding the L3 simulation later in this Section). A similar effect
can be seen at $t = 2.9$~Myr in the L4T1 simulation. The outflow
itself stalls and begins collapsing somewhat later, resulting in peaks
in the fragmentation rate: at $t = 2.9$~Myr for L3T1, $t = 4$~Myr for
L4T1 and $t = 4.8$~Myr for L5T1. It is worth noting that both the mean
and the peak fragmentation rates in the L5T1 simulation are lower than
in the L3T1 and L4T1 simulations. This happens because in L5T1, the
AGN is powerful enough to remove some of the gas to very large
distances, so there is less gas available for star formation.  This
effect is even clearer in L10T1, which shows a fragmentation rate
rather similar to the control simulation.

The effect of AGN activity upon the gas distribution can be understood
by considering the phase diagrams of the gas. In Figure
\ref{fig:phaseplots}, we show the density-temperature plots of the gas
in the L3T1 simulation at $t = 0.75$~Myr, $1.75$~Myr and $2.75$~Myr
(left, middle and right panels, respectively). At early times, a lot
of gas has cooled down to the temperature floor (the dashed line) and
is able to fragment. AGN activity pushes the gas away from this floor
by heating it. The heating-cooling prescription of
\citet{Sazonov2005MNRAS} that we use allows gas to be in approximate
thermal equilibrium at two temperatures, with rather weak dependence
on density $T \simeq 2 \times 10^4$~K and $T \simeq 10^5$~K. Both of
these temperatures are much higher than the temperature floor for all
except the densest gas, therefore gas can fragment into resolved
clumps with masses $M_{\rm J} \sim 10^5 n_4^{-1/2} \msun$, where $n_4
\equiv n/10^4$~cm$^{-3}$. Clumps of this mass and density have
electron scattering optical depth $\tau_{\rm es} \sim 0.1$, therefore
the assumption that the gas is optically thin to the radiation field
is valid (but see Section \ref{sec:shielding} for discussion of this
assumption in general). This heated gas, as a result, is kept from
fragmenting and forming stars. The gas is also compressed, but that
effect is not powerful enough to compensate for the heating and
therefore fragmentation ceases. Once the AGN switches off, gas can
cool down, and fragment again, but typically cools down while being
compressed by the surrounding hot gas, leading to an increase in
density.

Beyond $t = 5$~Myr, simulations L0.5T1, L1T1, L2T1 and, to some
extent, L3T1 and L4T1 enter a different evolutionary phase. By this
time, the initial turbulence has decayed significantly and material is
falling toward the centre in dense filaments. These filaments reach
the threshold density for fragmentation and therefore produce a lot of
sink particles, with fragmentation rates rising to $>150
\msun$~yr$^{-1}$ for a short while. These peaks are not realistic,
because in a real system, stellar feedback would prevent such coherent
filaments from maintaining these high densities.

The mean fragmentation rate in the period between $t=2$ and $5$~Myr
($6$~Myr for L5T1, to encompass the peak at $4.8$~Myr) is
$\dot{M}_{\rm mean} = 10, 11, 16, 37, 47, 26$ and $15 \msun$~yr$^{-1}$
for L0.5T1, L1T1, L2T1, L3T1, L4T1, L5T1 and L10T1, respectively. In
the control simulation over the same period, $\dot{M}_{\rm mean, c} =
10 \msun$~yr$^{-1}$. We see that the L0.5T1, L1T1 and L2T1 simulations
do not show significant enhancement of mean fragmentation rate over
the control simulation, while L3T1, L4T1 and L5T1 do; L10T1, once
again, shows a fragmentation rate similar to that of the control
simulation. In particular, L4T1 shows a mean fragmentation rate which
is nearly five times greater than that in the control simulation. We
interpret this result as a sign that there is a critical luminosity
which leads to the highest post-AGN star formation rate. At
luminosities higher than critical, the AGN outflow removes some of the
gas from the galaxy, reducing the amount of fuel available for star
formation. At luminosities lower than critical, the outflow is not as
efficient in compressing the gas to high densities and creating
conditions favorable to star formation The precise value of the
critical luminosity certainly depends on the properties of the gas
distribution and the galactic potential, but exploration of these
parameters is beyond the scope of the current paper.

In Figure \ref{fig:fragrates_inf}, we show fragmentation rates of gas
in five simulations with continuous AGN activity; we do not show or
analyse simulation L0.5, since its fragmentation rate are either
identical to the control simulation, or simulations L5 and L10, since
their fragmentation rates are negligible, after $t=2$~Myr. Simulations
L1 and L2 do not experience significant fragmentation after a short
burst between $t = 1$ and $2$~Myr until $t > 5$~Myr. This happens
because AGN heating keeps the gas above the temperature floor, even
though the outflow is unable to remove the gas from the galaxy (see
also Figure \ref{fig:phaseplots}). After $t = 5$~Myr, once turbulence
has decayed, the AGN is unable to heat up the dense infalling
filaments and a burst of fragmentation occurs as in the control
simulation. Simulation L4 quite predictably shows very little
fragmentation, as the AGN outflow efficiently removes gas out of the
potential well; even though this gas cools, it never reaches high
enough densities for fragmentation. The L3 simulation is most
interesting, showing three peaks of fragmentation. In this simulation,
the gas undergoes a cycle of expulsion to large distances, cooling,
recollapse toward the centre and significant fragmentation, then
heating and expulsion again as the densest gas is turned into sink
particles. After three such cycles, the outflowing shell is no longer
dense enough to cool and collapse efficiently.

\subsection{Locations of fragmentation}

\begin{figure*}
  \centering
    \includegraphics[trim = 0 0 0 0, clip, width=0.49 \textwidth]{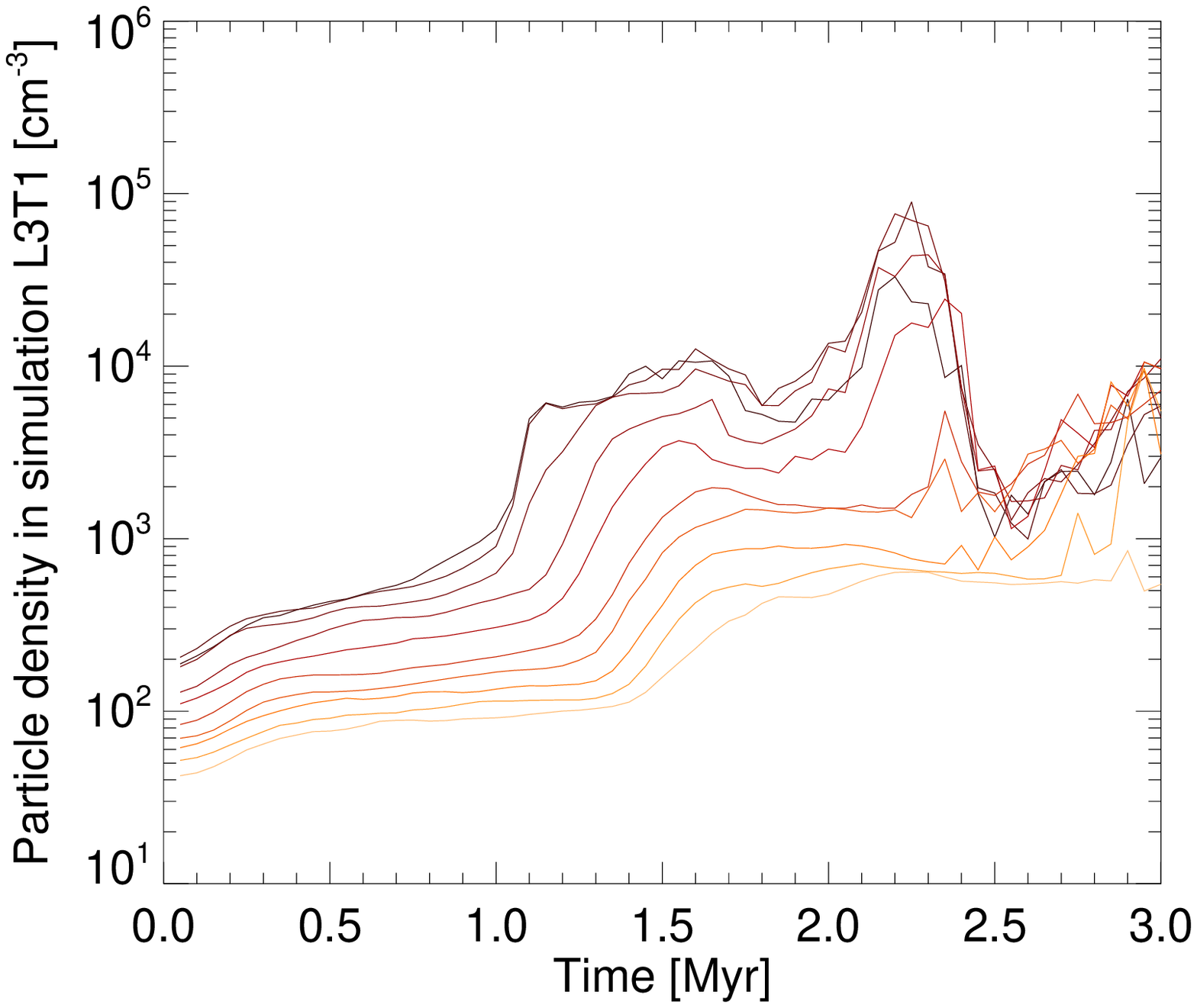}
    \includegraphics[trim = 0 0 0 0, clip, width=0.49 \textwidth]{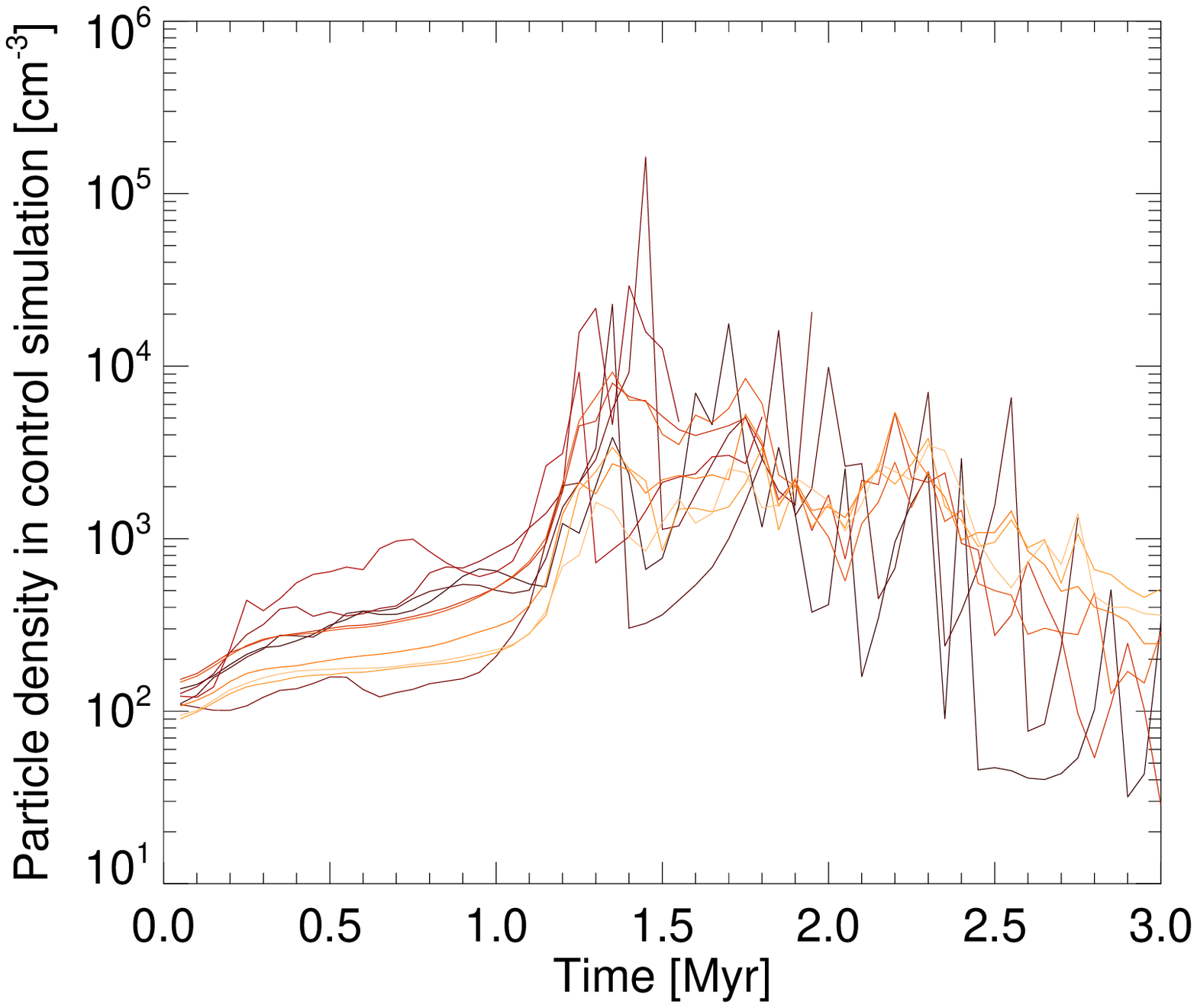}
  \caption{Averaged histories of particle densities as function of
    time for 10 groups of particles. {\it Left}: particles which turn
    into sink particles at $1$~Myr $< t < 3$~Myr in the control
    simulation, but remain as gas particles for at least $3$~Myr in
    the simulation L3T1, grouped by fragmentation time in the control
    simulation, in intervals of 0.2 Myr, with $t_{\rm frag}$
    increasing from darkest to brightest lines. {\it Right}: particles
    which turn into sink particles at $1$~Myr $< t < 3$~Myr in the
    simulation L3T1, but remain as gas particles for at least $3$~Myr
    in the control simulation, grouped by fragmentation time in
    simulation L3T1, in intervals of 0.2 Myr, with $t_{\rm frag}$
    increasing from darkest to brightest lines.}
  \label{fig:denshist}
\end{figure*}

In order to get a better understanding of the properties of the gas
which turns into sink particles, we consider the locations where sink
particles appear. We choose two representative simulations, L3T1 and
L5T1. For each sink particle that appears in those simulations during
the first $3$~Myr of evolution, we check the location where it
appeared and consider its properties at the moment of formation.

Figure \ref{fig:fragloc} shows the radial coordinates where sink
particles appeared in simulations L3T1 (left) and L5T1 (right),
separated into three time bins: particles appearing before the AGN
switches on at $t = 1$~Myr, particles appearing during the period of
AGN activity ($1$~Myr $< t_{\rm frag} < 2$~Myr) and particles
appearing after the AGN switches off ($t_{\rm frag} > 2$~Myr).

As expected, fragmentation before the AGN switches on is confined to
the inner edge of the shell, located at $R = 0.1$~kpc (blue dashed
line). During the period of AGN activity, there is very little
fragmentation (green dot-dashed line), and it happens slightly further
outward than before AGN activity, as the gas is being pushed away by
the outflow. Once the AGN switches off, fragmentation rates increase
significantly. In the L3T1 simulation, a two-peaked radial
distribution emerges (red triple-dot-dashed line). Some of the gas
falls in toward the SMBH and fragments as its density increases close
to the accretion boundary at $R = 0.01$~kpc. However, a significant
amount of fragmentation happens at the outer edge of the outflow
bubble at $0.3$~kpc $< R < 0.6$~kpc. This shows that the outflow
compresses some gas and induces fragmentation there. This effect is
even more evident in L5T1, where most of the fragmentation after $t =
2$~Myr happens in a shell with $0.4$~kpc $< R < 0.8$~kpc.

Figure \ref{fig:fragvel} highlights the radial velocities of sink
particles at the moment of formation and compares them with the radial
velocities of gas at $t = 3$~Myr in the two simulations. The general
trend in both L3T1 (left panel) and L5T1 (right panel) is similar:
fragmenting gas has significantly higher mean radial velocity than all
gas. The difference in velocities can be as large as $200$~km/s,
especially at $R < 0.03$~kpc and $R \sim 0.1-0.2$~kpc in L3T1 and $R
\sim 0.05-0.09$~kpc in L5T1. Typically fragmenting gas has a $\sim
50-100$~km/s higher mean radial velocity than all gas. This behaviour is
expected, because a lot of fragmentation occurs at the edges of the
outflowing bubbles, rather than in the infalling filaments.

It is rather interesting that the radial velocity difference is so
large at low radii in L3T1. This happens because gas that is falling
in radially toward the centre, with large negative radial velocity,
gets stretched out and its density decreases, preventing
fragmentation. However, the gas that has some angular momentum and
begins orbiting the SMBH accumulates and fragments into sink
particles. Therefore the sink particles tend to have very low absolute
radial velocities, but rather high tangential velocities, in the
central regions of the simulation. This is another effect of the AGN:
it helps redistribute the angular momentum of gas, providing avenues
for parcels of gas with nonzero angular momentum to fall close to the
centre and begin orbiting the SMBH, rather than falling in after
losing angular momentum due to dynamical interactions with the rest of
the gas \citep[this effect has been analysed before
  by][]{Dehnen2013ApJ}.

One more point to note is that the stars forming at the edge of the
outflow bubble inherit its expansion velocity, and therefore some of
them have radial velocities of several hundred km/s. Such stars would
be distinguishable by their kinematics, and may provide evidence of
recent AGN activity in the host galaxy. We return to this point in
Section \ref{sec:kinematics}.

\subsection{AGN effects on individual gas clumps}

We have shown that AGN activity has a global effect on fragmentation
rates in the host galaxy spheroid. However, this global effect can be
achieved in several ways. For example, the AGN outflow may push
already dense gas clumps above the density threshold. Alternatively,
outflow compression might be strong enough to produce new clumps which
would otherwise disperse or not form at all. In order to distinguish
between these possibilities, we consider the properties and evolution
of three related subsets of particles:

\begin{itemize}
\item Those gas particles which turn into sink particles between $t =
  1$~Myr and $t = 3$~Myr in the control simulation;
  
\item Those gas particles which turn into sink particles between $t =
  1$~Myr and $t = 3$~Myr in the simulation L3T1;

\item Those gas particles which turn into sink particles between $t =
  1$~Myr and $t = 3$~Myr in the simulation L5T1.
\end{itemize}

We track the first subset of particles in all three simulations -
control, L3T1 and L5T1. By considering how the particles that would
have turned into sink particles without AGN activity evolve in the
presence of an AGN outflow, we can quantify the quenching effect of
AGN activity on star formation. The other two subsets are tracked in
their respective simulations and in the control simulation. This way,
we can quantify the enhancing effect of AGN activity.

In the control simulation, we select 5113 particles. Of these, only
182 turn into sink particles in the L3T1 simulation. The rest are
either accreted by the SMBH (1648 particles) or remain in the gas
phase by $t = 3$~Myr (3283 particles). Conversely, of the 7020
particles selected in the L3T1 simulation, 5952 are accreted in the
control simulation and 886 remain in the gas phase by $t = 3$~Myr,
with the remaining 182 being the particles that turn into sink
particles in both L3T1 and control simulations.

Next, we consider the density histories of the particles that turn
into sink particles in one of the simulations, but remain gaseous in
another. Figure \ref{fig:denshist} shows the density evolution of
these particles. On the left, we represent the 3283 particles which
turn into sink particles in the control simulation, but remain gaseous
by $t = 3$~Myr in the L3T1 simulation, by grouping them into ten
subsets by time of fragmentation in the control simulation and
averaging the density histories in each subset. The line colours, from
darkest to brightest, indicate increasing $t_{\rm frag}$ from $1$~Myr
to $3$~Myr in intervals of $0.2$~Myr. On the right, we represent, in
analogous fashion, the 886 particles which turn into sink particles in
L3T1, but not in the control simulation. For the particles selected
from the control simulation (left panel), there is a trend of
initially increasing density, which jumps up and plateaus soon after
$t = 1$~Myr. In the control simulation, these particles experience
further density increase and turn into sink particles, but AGN
feedback prevents them from doing so in L3T1. Later, once the AGN
activity ceases, the sudden reduction in heating rate results in
further contraction of the densest gas and increase in their
densities. However, after $\sim 2.5$~Myr, a lot of dense clumps
evaporate in the outflow bubbles and the mean density of the
constituent particles decreases significantly. The particles which
fragment at late times in the control simulation (brightest lines)
experience a rather gradual density increase throughout the period
until $t = 3$~Myr, as they are further from the AGN and/or better
shielded from the outflow by dense filaments.

Among the particles selected from simulation L3T1, only a single trend
is evident independently of fragmentation time. In simulation L3T1,
these particles enter dense filaments that get further compressed by
the outflow bubbles, leading to fragmentation. In the control
simulation, the filaments are not compressed as much, so there is less
fragmentation there. At first (up to $t \sim 1.5$~Myr), the formation
and contraction of filaments is the dominant process, so the particle
density increases. At later times, stretching of the filaments starts
to dominate, leading to a gradual decrease in density.

The difference between the control simulation and simulation L5T1 is
even more striking: only 76 particles turn into sink particles between
$t = 1$ and $3$~Myr in both simulations. 460 particles that turn into
sinks in the control simulation are accreted by the SMBH in L5T1,
while 4577 remain in the gaseous state. 2903 particles that turn into
sinks in L5T1 are accreted in the control simulation, and 708 remain
gaseous. The density histories look qualitatively similar to those
depicted in Figure \ref{fig:denshist}.

Overall, these results suggest that AGN activity has a very powerful
effect on the gas distribution, but that effect is neither purely
quenching star formation, nor purely enhancing. Individual dense gas
clumps, with densities of a few times $10^4$~cm$^{-3}$, can survive
for some time in AGN outflows, but are eventually dispersed. Their
fragmentation can be suppressed by the AGN radiation field. On the
other hand, AGN outflows push together material which can then cool
down and form new dense clumps which experience significant
fragmentation and subsequent star formation. In other words, the
effect of AGN activity upon star formation in its host galaxy is more
global (changing which has forms stars) than local (affecting
pre-existing star formation regions), even though local effects are
not completely negligible either.

\subsection{Effects of self-shielding} \label{sec:shielding}

\begin{figure}
  \centering
    \includegraphics[trim = 0 0 0 0, clip, width=0.49 \textwidth]{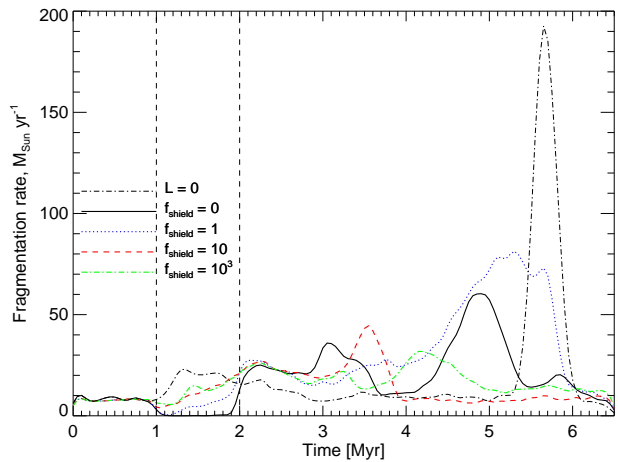}
  \caption{Fragmentation rates in the four L5T1 simulations with
    different levels of self-shielding (thick lines) plus the control
    simulation without AGN activity (thin dashed line). Line labels
    give the value of $f_{\rm shield}$, a dimensionless factor that
    increases the gas optical depth above its locally-computed value;
    see eq. \ref{eq:optdepth}.}
  \label{fig:fragrates_shield}
\end{figure}

So far, we have assumed that all the gas in the simulations is
optically thin to the AGN radiation field. This is not generally true,
as a column density of $N_{\rm H} \simeq 1.7 \times 10^{24}$~cm$^{-2}$
is enough to make the gas Compton-thick. Such large column densities
might not be reached on large scales \citep[although some $10-20\%$ of
  AGN in the local Universe might be Compton-thick;
  cf. ][]{Akylas2012A&A}, but individual clumps might be strongly
shielded. A detailed analysis of this shielding is beyond the scope of
the paper, but we can estimate the impact of gas self-shielding by
reducing the ionisation parameter of the gas assuming the gas has an
optical depth proportional to the local density and distance from the
AGN, as described in equation (\ref{eq:optdepth}).

Figure \ref{fig:fragrates_shield} shows the fragmentation rates of
simulations L5T1, L5T1A1, L5T1A10 and L5T1A1000, with progressively
stronger shielding, in addition to the control simulation with $L =
0$. Considering the time period $1 < t/{\rm Myr} < 2$, even $f_{\rm
  shield} = 1$ allows some fragmentation to occur, while $f_{\rm
  shield} = 10$ and $f_{\rm shield} = 1000$ result in almost identical
fragmentation rates, both $\sim 2$ times lower than that in the
control simulation. From this we can conclude that both the kinetic
power of AGN outflows and the AGN radiation field reduce the
fragmentation rate by similar amounts, so that when the radiation
field effect is negated by shielding, the fragmentation rate drops
only by $50\%$.

After the AGN switches off, all four simulations evolve in a similar
manner at first, with statistically insignificant differences up to $t
\sim 4$~Myr. At that point, the efficiently shielded simulations
($f_{\rm shield} = 10$ and $f_{\rm shield} = 1000$) continue to have
fragmentation rates not much higher than those in the control
simulation, while the less shielded ones experience bursts of
fragmentation. This seemingly paradoxical outcome can be understood as
a result of isobaric gas cooling. When gas self-shielding is low or
nonexistent, more gas heats up to intermediate temperatures
$10^5-10^6$~K during the AGN phase. When the AGN switches off, this
gas cools down while still being compressed by the hottest material
with $T > 10^7$~K. Since the hottest material cools down
inefficiently, the cooling of the intermediate-temperature gas is
mostly isobaric, i.e. its density increases as temperature
decreases. In the unshielded simulations, this results in a lot of gas
cooling down to the temperature floor at very high density and rather
large temperature $T > 10^4$~K. On the other hand, in the shielded
simulations, a lot of gas piles up at $T \simeq 10^4$~K without
reaching the temperature floor first. Therefore, the unshielded
simulations have more material available for fragmentation, resulting
in a higher fragmentation rate.

The true effect of gas self-shielding is likely to be somwhere between
the extremes explored here. Diffuse gas, especially that within
outflows, is most likely less shielded than assumed here, while dense
gas might be shielded better. Long filaments, which tend to form
especially in lower-luminosity simulations, also shield gas behind
them, potentially increasing the fragmentation rate during the AGN
phase, but reducing it afterward. We plan to explore these issues in
the future (see also Section \ref{sec:improv}).

\section{Discussion and conclusions} \label{sec:discuss}

\subsection{Summary of results}

In this paper, we presented results of a number of simulations
designed to showcase the effects that AGN activity has upon
fragmentation rates in a surrounding turbulent gas shell. The
idealised initial conditions can mimic a gas-rich galaxy bulge. We ran
simulations with five values of AGN luminosity (equal to Eddington
luminosity for a $1-5 \times 10^8 \msun$ SMBH) and two values of
duration of activity (1 Myr and infinity), plus a control simulation,
as well as considering the effects of gas self-shielding against the
AGN radiation field. The main results are:

\begin{itemize}
\item AGN activity efficiently stops gas fragmentation while the AGN
  is active, so long as the AGN is powerful enough. In the two
  lowest-luminosity simulations, fragmentation resumes before the AGN
  switches off, even though those luminosities are formally large
  enough to drive a large-scale outflow through the gas of density
  equal to the mean density in the simulations.

\item Once the AGN switches off, fragmentation resumes very quickly,
  as gas compressed by the outflowing bubbles cools down.

\item There is a critical AGN luminosity which produces a maximum
  enhancement of fragmentation rate. For our setup, this luminosity is
  $L_{\rm cr} \simeq 5 \times 10^{46}$~erg~s$^{-1}$, approximately
  four times higher than the formal luminosity necessary for the AGN
  momentum to drive a large-scale outflow. At this critical
  luminosity, the fragmentation rate is enhanced by almost a factor 5.

\item In the gas shell affected by the AGN outflow, fragmentation
  occurs in two regions: dense filaments as they fall in toward the
  SMBH and the edges of the outflow bubbles. The second region becomes
  progressively more important with increasing AGN luminosity.

\item Fragmenting gas has systematically higher radial velocities than
  all gas. In particular, at the edge of the AGN bubbles, fragments
  form with positive radial velocities and may be expected to move far
  out from their locations of formation.

\item Most of the fragmenting particles are different between the
  control simulation and simulations with AGN. This shows that AGN
  outflows tend to destroy the majority of dense gas clumps, but later
  produce conditions favourable for the formation of new dense clumps.
  
\end{itemize}

\subsection{Kinematics of young stars} \label{sec:kinematics}

The gas that fragments in simulations with AGN, and the sink particles
themselves (and hence the stars they represent) move with
systematically higher radial velocities than the mean velocities of
the gas. This is not unexpected: sink particles appear preferentially
on the outskirts of outflow bubbles and inherit the velocity of those
bubbles, which keep expanding for a long time \citep[up to an order of
  magnitude longer than the AGN phase itself, c.f. ][]{King2011MNRAS}
after the AGN has switched off. In simulations where the outflow
bubbles can escape from the galaxy (L4T1, L5T1 and L10T1), sink
particles forming among them have almost only positive radial
velocities, with $\sim 5-10\%$ of all sink particles forming with
$v_{\rm r} > \sigma$. This leads to two consequences for observing
such systems:

\begin{itemize}

\item The stellar population in the spheroid of a galaxy that recently
  underwent a period of AGN activity has a velocity structure biased
  toward positive radial velocities.

\item Some newborn stars might move to the halo of the host galaxy or
  escape from it altogether.

\end{itemize}

The importance of these two effects depends strongly on the gas
content of the galaxy spheroid at the time the outflow is inflated. In
gas-poor galaxies, the outflow propagates rapidly, but gas density
does not reach large enough values for fragmentation. In gas-rich
galaxies, star formation is rapid, but the outflow might not attain
large enough radial velocities to produce outflowing stars. We plan to
investigate this dependence in future work (see also Section
\ref{sec:improv}).

A radial anisotropy of stellar velocity dispersion would be difficult
to identify and to interpret as evidence of recent AGN activity due to
difficulties in modelling stellar orbits and their anisotropies in
external galaxies \citep{Cappellari2013MNRAS,
  Falcon2016ASSL}. Nevertheless, we predict that galaxies showing
evidence of recent AGN activity, such as hot gas bubbles on galactic
outskirts \citep[analogous to the Fermi bubbles,][]{Su2010ApJ} or
young stars very close to the nucleus \citep[analogous to the disc of
  young stars in the Galactic centre,][]{Paumard2006ApJ} may also have
radially-biased stellar velocity dispersions. By itself, this feature
would not distinguish galaxies which recently underwent an AGN phase,
because radially-biased velocities may appear for other
reasons. However, in combination with other lines of evidence,
information about stellar velocity dispersion would strengthen the
case for recent AGN activity. A more detailed analysis, showing higher
radial velocities for young stars than for old ones, would strengthen
this argument further.

Several authors predicted the possibility of high-velocity stars
forming due to AGN activity. For example, \citet{Silk2012A&A} proposed
that AGN jets can induce star formation and lead to formation of
hypervelocity stars. In \citet{Zubovas2013MNRAS}, we also predicted
that stars forming in AGN outflows might escape galaxies, using a more
idealised spherical initial gas distribution than here. In particular,
this process would produce high-metallicity stars in galactic
outskirts, which may be used as another piece of evidence for recent
AGN activity. Although identifying individual stars may be difficult,
stacking observations can help reveal their influence upon
extragalactic light \citep{Zibetti2005MNRAS}.

It is interesting to note that recently, star formation has been
detected within a galactic outflow. Observations of a merging system
IRAS F23128-5919 revealed a spatially-resolved outflow with clear
evidence of star formation occurring within, at rates of $\sim 15
\msun$~yr$^{-1}$ \citep{Maiolino2017Natur}. More such
systems may be discovered in the near future, and exploration of the
correlations between AGN properties and star formation rates in their
outflows will provide new insights into AGN-galaxy coevolution.

Finally, some of the stars ejected from the bulge of a disc galaxy by
this mechanism may not be able to escape from the galaxy altogether,
but might end up falling on to the galaxy disc, contributing to its
stellar population. The numbers of such stars are unlikely to be
large, but they might be discernible due to having abnormally large
metallicity compared with the average for their present location.

\subsection{Implications for interpretation of outflow observations}

It is generally thought that starbursts precede AGN activity by a
period of $\sim 10^8$~yr \citep{Davies2007ApJ, Schawinski2009ApJ,
  Wild2010MNRAS}. Here we show that the opposite might also be true: a
starburst may be seen in a galaxy for several Myr after the AGN phase
ends. Such starbursts would occur in the spheroidal component of the
galaxy \citep[although the disc might experience a starburst of its
  own due to the outflow compressing its gas vertically; see
][]{Zubovas2013MNRASb, Zubovas2016MNRASb}. They might be accompanied
by the presence of a fading AGN, which is no longer able to prevent
gas from collapsing and fragmenting. We note that AGN are sometimes
associated with elevated star formation both throughout the host
galaxy \citep{Santini2012A&A} and in the central regions
\citep{LaMassa2013ApJ}, but determining the causal connection between
the two processes may be very difficult. There are several ways to
disentangle this relationship:

\begin{itemize}

\item In this paper, we show that SFR enhancement should be correlated
  with the fading of the AGN, so detection of a fading AGN together
  with a starburst in the central region of the galaxy would suggest
  that the AGN has enhanced the host galaxy's SFR.

\item Observations with high spatial resolution would help determine
  whether the starburst is spatially correlated with the AGN
  outflow. For example, \citet{Carniani2016A&A} present observations
  of high-z AGN outflows, showing some evidence that star formation
  might be enhanced along the edges of the outflow bubbles, consistent
  with the results of our simulations.

\item Small-scale numerical simulations of the star formation process
  in AGN-outflow-affected gas might help distinguish this mode of star
  formation from regular star formation. For example, AGN-enhanced
  star formation would have higher gas densities in the star-forming
  regions: in our simulations, sink particles that appear after the
  AGN switches off form from $\sim 10^2$ times denser gas on average
  than sink particles that appear before the AGN switches on. A more
  detailed analysis of these aspects is left for a future study.

\end{itemize}

\subsection{Drawbacks and possible improvements} \label{sec:improv}

While our simulations reveal interesting behaviour, they are idealised
in many ways, limiting their usefulness in directly predicting
observational signatures of AGN-influenced star formation. We plan to
address these issues in future publications. Here we briefly comment
on the major drawbacks of the current simulations.

The complex evolution of AGN outflows and their interaction with the
ISM depend significantly on the distribution of the ISM in phase
space. Therefore, more realistic initial conditions are necessary in
order to be able to predict AGN outflow effects upon star formation in
various galaxies. We plan to run similar simulations for galaxies with
bulges with different gas content, size and shape, to estimate the
importance of AGN-induced star formation at different redshifts
(represented by gas fraction) and for different galaxy morphological
types (represented by bulge size and shape).

As discussed above, in Section \ref{sec:shielding}, gas self-shielding
may have a significant impact upon gas fragmentation rates. Here we
estimated its effect by assuming that gas density is never very
different from an isothermal distribution, which is clearly incorrect,
especially with high-luminosity AGN driving a large outflow. A more
realistic estimate of the optical depth is required. A full radiative
transfer implementation would make the simulation unfeasibly slow, but
using a separate set of virtual particles to represent photon packets
of the AGN radiation field, in addition to the virtual particles
representing the relativistic wind, may solve the issue. We are
currently working on developing such a feedback prescription (Sabulis
\& Zubovas, in preparation) and will use it to update AGN outflow
simulations when it is ready. This improvement might also help
understand how star formation occurs while the AGN is still active and
heating the gas.

Small-scale simulations of individual clumps evolving within a hotter
AGN outflow would help understand the star formation process
better. In idealised simulations, we showed that fragmentation is more
efficient in externally-compressed clouds \citep{Zubovas2014MNRASc},
but those simulations did not include stellar feedback. Feedback, in
the form of radiation and winds from newborn stars, might heat dense
clumps and perhaps break them up, reducing the numbers and masses of
stars forming there. Therefore, including stellar feedback in both
small-scale and larger simulations of AGN-induced star formation is an
important step.

Finally, our current simulations had no connection between the
accretion on to the SMBH particle and the AGN luminosity. While it is
unlikely that such a connection would exist on the short (several Myr)
timescales probed here, it matter for longer-term simulations. AGN
activity regulates the SMBH mass supply; it is important to understand
how this self-regulation is affected by star formation within the AGN
outflow.

The interaction between AGN outflows and star formation in the host
galaxy is a complex issue, involving many spatial and temporal
scales. The possibility of star formation within the outflows
themselves is a rather new aspect of the problem, and our simulations
are one of the first analysing {\it when and where} this star
formation might occur. In this way, our work complements others in the
field of AGN-galaxy coevolution.

\section*{Acknowledgments} 
KZ is funded by the Research Council Lithuania through the National
Science Programme grant no. LAT-09/2016. MAB acknowledges support by
the ERC starting grant 638707 ``BHs and their host galaxies:
co-evolution across cosmic time.'' Simulations were performed on
resources at the High Performance Computing Center “HPC Sauletekis” in
Vilnius University Faculty of Physics.

%\bibliographystyle{mnras}
%\bibliography{../../../zubovas} \label{lastpage}

\end{document}